\documentclass[nopreprintline]{elsarticle}

\usepackage[T1]{fontenc}

\usepackage[margin=2.25cm]{geometry}
\usepackage{amsmath}
\usepackage{amssymb}
\usepackage{cases}
\usepackage{empheq}
\usepackage{bm}

\usepackage[per-mode=symbol,
            inter-unit-product=\cdot,
            separate-uncertainty=true,
            print-unity-mantissa=false,
            table-alignment-mode=none,
            table-alignment=right,
            exponent-product=\cdot]{siunitx}
\usepackage{physics}
\AtBeginDocument{\RenewCommandCopy\qty\SI}
\ExplSyntaxOn
\msg_redirect_name:nnn { siunitx } { physics-pkg } { none }
\ExplSyntaxOff

\usepackage{pifont}

\usepackage[hypertexnames=false,
            colorlinks=true,
            pdfborderstyle={/S/U/W 0},
            citecolor=blue]{hyperref}
\usepackage[capitalize]{cleveref}

\usepackage{graphicx} 
\usepackage{subcaption}
\usepackage{xcolor}

\usepackage{booktabs}
\usepackage{listings}
\usepackage{makecell}
\usepackage{multirow}
\usepackage{pdflscape} 

\usepackage{todonotes}

\everymath{\displaystyle}

\definecolor{lifex}{HTML}{f60248}
\definecolor{codegreen}{rgb}{0,0.6,0}
\definecolor{codegray}{rgb}{0.5,0.5,0.5}
\definecolor{codepurple}{rgb}{0.58,0,0.82}
\definecolor{backcolor}{rgb}{0.95,0.95,0.92}
\colorlet{linkcolor}{blue!50!black}

\lstdefinestyle{mystyle}{
    backgroundcolor=\color{backcolor},
    commentstyle=\color{codegreen},
    keywordstyle=\color{lifex},
    numberstyle=\tiny\color{codegray},
    stringstyle=\color{codepurple},
    basicstyle=\ttfamily\footnotesize,
    breakatwhitespace=false,
    breaklines=true,
    captionpos=b,
    keepspaces=true,
    numbers=none,
    numbersep=5pt,
    showspaces=false,
    showstringspaces=false,
    showtabs=false,
    tabsize=4,
    frame=single,
}

\lstdefinelanguage{prm}{
    keywords={subsection, set, end},
    comment=[l]{\#}
}

\lstdefinelanguage{xml}{
    keywords={value, default\_value, documentation, pattern, pattern\_description}
}

\lstdefinelanguage{json}{
    keywords={value, default\_value, documentation, pattern, pattern\_description}
}

\newcommand{\documentationurl}{https://lifex.gitlab.io/lifex-cfd/}

\renewcommand{\k}{\mathrm{k}}
\renewcommand{\laplacian}{\bm\Delta}
\newcommand{\lifex}{\texttt{life\textsuperscript{\color{lifex}{x}}}}
\newcommand{\lifexcfd}{\texttt{\lifex{}-cfd}}
\newcommand{\dealii}{\texttt{deal.II}}
\newcommand{\sigmak}{\Sigma_\mathrm{k}}
\newcommand{\xhat}{\widehat{\bm x}}
\newcommand{\gammad}{\Gamma_t^\mathrm{D}}
\newcommand{\gamman}{\Gamma_t^\mathrm{N}}
\newcommand{\gammar}{\Gamma_t^\mathrm{R}}
\newcommand{\dt}{{\Delta t}}
\newcommand{\bdf}{\mathrm{BDF}}

\newcommand{\vel}{\bm u}
\newcommand{\displ}{\widehat{\bm d}}
\newcommand{\uale}{\bm u^\mathrm{ALE}}
\newcommand{\resistive}{\bm{\mathcal R}(\vel, \uale)}

\newcommand{\spacevelocity}{\bm{\mathcal V}_{\bm g}^t}
\newcommand{\spacepressure}{\mathcal Q^t}
\newcommand{\spacevelocityh}{\bm{\mathcal V}_{\bm g,h}^{t}}
\newcommand{\spacemomentumtesth}{\bm{\mathcal V}_{\bm 0}^h}
\newcommand{\spacepressureh}{\mathcal Q^t_h}

\newcommand{\velh}{\vel_h}
\newcommand{\ph}{p_h}

\newcommand{\velhcurrent}{\velh^{n+1}}
\newcommand{\phcurrent}{\ph^{n+1}}

\newcommand{\velhbdf}{\velh^{\mathrm{BDF}}}
\newcommand{\velhext}{\velh^{\mathrm{EXT}}}
\newcommand{\phext}{\ph^{\mathrm{EXT}}}

\newcommand{\ustar}{\velh^*}
\newcommand{\ualehcurrent}{\vel_h^{\mathrm{ALE}, n+1}}

\newcommand{\vh}{\bm v_h}
\newcommand{\qh}{q_h}

\DeclareSIUnit\mmhg{mmHg}

\usepackage{placeins}
\newdefinition{remark}{Remark}

\biboptions{sort&compress}
\begin{document}

\begin{frontmatter}

\title{lifex-cfd: an open-source computational fluid dynamics solver for cardiovascular applications}
\author[1,2]{Pasquale Claudio Africa\texorpdfstring{\corref{cor1}}{}}
\author[2]{Ivan Fumagalli}
\author[2]{Michele Bucelli}
\author[3,2]{\texorpdfstring{\\}{}Alberto Zingaro}
\author[2]{Marco Fedele}
\author[2]{Luca Dede'}
\author[2,4]{Alfio Quarteroni}
\address[1]{SISSA, International School for Advanced Studies, Mathematics Area, mathLab, Via Bonomea 265, 34136, Trieste, Italy}
\address[2]{MOX, Laboratory of Modeling and Scientific Computing, Dipartimento di Matematica, Politecnico di Milano, Piazza Leonardo da Vinci 32, 20133, Milano, Italy}
\address[3]{ELEM Biotech S.L., Pier01, Palau de Mar, Plaça Pau Vila, 1, 08003, Barcelona, Spain}
\address[4]{Institute of Mathematics, \'Ecole Polytechnique F\'ed\'erale de Lausanne, Station 8, Av. Piccard, CH-1015 Lausanne, Switzerland (Professor Emeritus)}
\cortext[cor1]{Corresponding author. E-mail: pafrica@sissa.it}

\begin{abstract}
Computational fluid dynamics (CFD) is an important tool for the simulation of the cardiovascular function and dysfunction. Due to the complexity of the anatomy, the transitional regime of blood flow in the heart, and the strong mutual influence between the flow and the physical processes involved in the heart function, the development of accurate and efficient CFD solvers for cardiovascular flows is still a challenging task. In this paper we present \lifexcfd{}, an open-source CFD solver for cardiovascular simulations based on the \lifex{} finite element library, written in modern C++ and exploiting distributed memory parallelism. We model blood flow in both physiological and pathological conditions via the incompressible Navier-Stokes equations, accounting for moving cardiac valves, moving domains, and transition-to-turbulence regimes. In this paper, we provide an overview of the underlying mathematical formulation, numerical discretization, implementation details and examples on how to use \lifexcfd{}.
We verify the code through rigorous convergence analyses, and we show its almost ideal parallel speedup. We demonstrate the accuracy and reliability of the numerical methods implemented through a series of idealized and patient-specific vascular and cardiac simulations, in different physiological flow regimes. The \lifexcfd{} source code is available under the LGPLv3 license, to ensure its accessibility and transparency to the scientific community, and to facilitate collaboration and further developments.
\end{abstract}


\end{frontmatter}

{\bf PROGRAM SUMMARY}

\begin{small}

\noindent
{\em Program Title:} \lifexcfd{} \\
{\em CPC Library link to program files:} (to be added by Technical Editor) \\
{\em Developer's repository link:} \url{https://gitlab.com/lifex/lifex-cfd} \\
{\em Code Ocean capsule:} (to be added by Technical Editor) \\
{\em Licensing provisions (please choose one):} \href{https://www.gnu.org/licenses/lgpl-3.0.html}{LGPLv3} \\
{\em Programming language:} C++ (standard $\geq 17$) \\
{\em Supplementary material:} \url{https://doi.org/10.5281/zenodo.7852088} contains the application executable in binary form, compatible with any recent enough \texttt{x86-64 Linux} system, assuming that \texttt{glibc} version $\geq$ \texttt{2.28} is installed. Data and parameter files necessary to replicate the test cases described in this manuscript are also available. \\
{\em Nature of problem:} the program allows to run computational fluid dynamics simulations of cardiovascular blood flows in physiological and pathological conditions, modeled through incompressible Navier-Stokes equations, including moving cardiac valves, moving domains (such as contracting cardiac chambers) in the arbitrary Lagrangian-Eulerian framework, and transition-to-turbulence flow. Given the scale of the typical applications, the program is designed for parallel execution. \\
{\em Solution method:} the equations are discretized using the Finite Element method, on either tetrahedral or hexahedral meshes. The software builds on top of \dealii{}, implementing the mathematical models and numerical methods specific for CFD cardiovascular simulations. Parallel execution exploits the MPI paradigm. The software supports both Trilinos and PETSc as linear algebra backends. \\
{\em Additional comments including restrictions and unusual features:} the program provides a general-purpose executable that can be used to run CFD simulations without having to access or modify the source code. The program allows to setup simulations through a user-friendly yet flexible interface, by means of readable and self-documenting parameter files. On top of that, more advanced users can modify the source code to implement more sophisticated test cases. \lifexcfd{} supports checkpointing, i.e. simulations can be stopped and restarted at a later time. \\

\end{small}

\section{Introduction}
\label{sec:introduction}

Computational Fluid Dynamics (CFD) simulations have emerged as a powerful tool for understanding the complex hemodynamics of the cardiovascular system and the mechanisms underlying pathological conditions \cite{formaggia2010cardiovascular,chnafa2016image,collia2019analysis,luraghi2018numerical,luraghi2020impact,goubergrits2022ct,karabelas2022global,kronborg2022computational, mittal2016computational,quarteroni2019mathematical,this2020pipeline,zingaro2021hemodynamics,viola2020fluid,viola2022fsei,santiago2018fully,viscardi2010comparative, faggiano2013helical, tagliabue2017fluid, tricerri2015fluid,fumagalli2023fluid,paliwal2021presence,mill2021sensitivity,morales2021deep, oks2022fluid, santiago2022design}. Hemodynamics simulations can provide insights into complex flow patterns \cite{domenichini2005three, domenichini2007combined, seo2013effect, seo2014effect, tagliabue2017complex, dede2021computational, masci2020proof,  zingaro2021hemodynamics, corti2022impact, di2021computational, barnafi2022multiscale}, pressure distributions, and wall stresses \cite{sacco2018left, vedula2016effect} that are difficult or impossible to measure in vivo. These insights can aid in the development of novel diagnostic tools and therapies \cite{brown2023patient,collia2019analysis,kung2021hemodynamics, lantz2021impact,rigatelli2021applications, oks2023effect}.
Indeed, cardiovascular diseases are a major global health concern, responsible for a significant number of deaths each year \cite{timmis2020european,virani2020heart}.

However, despite the significant progress made in the computational modeling of the cardiovascular system, the development of accurate and efficient CFD solvers for cardiovascular applications remains a challenging task \cite{updegrove2017simvascular}, due in part to the complex anatomy and physical processes involved, as well as to the need for high-performance computing resources required to capture the dynamics of thin structures (such as valves) \cite{mittal2016computational, de2016numerical, marom2015numerical, spuhler2020high, votta2013toward} and small turbulent scales \cite{chnafa2014image, nicoud2018large}.
Indeed, most of the computational software libraries currently available on the market and open source only address some of these challenges, and there is room for enhancement in the development of a single comprehensive tool that provides, in a computationally efficient way, accurate results in terms of anatomical detail, flow kinematics, and stress reconstruction.

In this paper, we present \lifexcfd{}, an open-source CFD solver for cardiovascular applications in moving geometries with immersed structures. It is based on \lifex{} \cite{africa2022flexible}, a flexible, high performance library for multiphysics, multiscale and multidomain finite element problems building on the \dealii{} \cite{dealII93} Finite Element (FE) core.

The scientific community has developed several open-source CFD software packages for cardiovascular simulations. The availability of the code under open-source licenses ensures accessibility and transparency to the wider scientific community, facilitating collaboration and further developments. Every available software package is characterized by the different needs it addresses and by its requirements, regarding the complexity of the simulation and the desired level of resolution, the size of the computational domain, and the availability of computational resources. Popular state-of-the-art methodologies for predictive simulation in cardiovascular health and disease are thoroughly reviewed in \cite{mittal2016computational, beyond_cfd2023}. FEBio provides rich capabilities for vascular flow simulations, differing from other CFD programs mainly by the use of fluid dilatation instead of pressure as a primary variable \cite{maas2012febio}. $\mathcal{C}Heart$ is a framework for multiphysics finite element simulations in biomedical research, including a CFD solver with advanced numerical features, an Arbitrary Lagrangian-Eulerian (ALE) formulation and the Streamline Upwind Petrov-Galerkin (SUPG) stabilization \cite{CHeart2016}. simVascular provides a complete computational framework, from the construction of an anatomic model to finite element simulation and postprocessing \cite{updegrove2017simvascular}, with the possibility to solve the incompressible Navier-Stokes equations in an arbitrary domain and specifically designed for cardiovascular simulations. IBAMR is an adaptive and distributed-memory parallel implementation of the Immersed Boundary Method (IBM) has also been used in hemodynamics applications \cite{griffith2013ibamr}. Oasis is a high-level/high-performance open source Navier–Stokes solver is written in Python using building blocks from the FEniCS project \cite{mortensen2015oasis}.

ExaDG is a high order Discontinuous Galerkin CFD solver based on \dealii{}, exploiting a matrix-free algorithm equipped with multigrid smoothers, with a focus on accuracy and extreme parallel performance \cite{ExaDG2020}. Lethe allows to solve the incompressible Navier–Stokes equations using high-order continuous Galerkin formulations on quadrilateral and hexahedral adaptive meshes, leveraging the \dealii{} library \cite{blais2020lethe}. General-purpose packages such as OpenFOAM \cite{chen2014openfoam,brenneisen2021sequential,lyras2021comparison,chen2023hemodynamic} and FEniCS \cite{fenicsx2022} have also been used for cardiovascular simulations. All the mentioned packages, except for $\mathcal{C}Heart$, are readily available to the community under open-source licenses.

\begin{landscape}
\begin{table}
    \scriptsize
    \centering
    \begin{tabular}{r|c|c|c|c|c|c}
        & \textbf{\lifexcfd{}} & \textbf{FEBio} \cite{maas2012febio} & \textbf{CHeart} \cite{CHeart2016} & \textbf{simVascular} \cite{updegrove2017simvascular} & \textbf{IBAMR} \cite{griffith2013ibamr} & \textbf{Oasis} \cite{mortensen2015oasis} \\\hline
        \textbf{Stabilization} & SUPG + PSPG & --- & SUPG & SUPG + PSPG & --- & --- \\\hline
        \textbf{Turbulence} & VMS-LES & --- & SUPG & --- & --- & LES, DNS \\\hline
        \textbf{Domain displacement} & ALE & ALE + FSI  & ALE & ALE + FSI  & Immersed Boundary & None \\\hline
        \textbf{Space discretization} & Finite Elements & Finite Elements & Finite Elements & Finite Elements & \makecell{Immersed Boundary\\on Cartesian grids} & Finite Elements \\\hline
        \makecell[r]{\textbf{Element types (degrees)}} & \makecell{Hexahedra (arbitrary)\\Tetrahedra (1, 2)} & \makecell{Hexahedra (1)\\Tetrahedra (1, 2)\\Pentahedra (1)} & \makecell{Tetrahedra (0-5)\\Hexahedra (0-5)\\Prisms (0-5)\\User-specified} & \makecell{Tetrahedra (1, 2)\\Hexagonal bricks (1, 2)\\Wedges} & --- & \makecell{As supported by FEniCS} \\\hline
        \makecell[r]{\textbf{Time stepping}} & BDF of order 1, 2, 3 & \makecell{Generalized-$\alpha$\\method} & Implicit/Explicit & \makecell{Generalized-$\alpha$\\method} & \makecell{Predictor-corrector\\Picard iteration} & Fractional step \\\hline
        \textbf{Linear solvers} & GMRES & \makecell{Pardiso\\FGMRES\\CG} & GMRES & Custom \cite{esmaily2013new,esmaily2015bi} & \makecell{Krylov solvers,\\PETSc} & BiCGStab \\\hline
        \textbf{Preconditioners} & \makecell[c]{aSIMPLE with\\AMG\\Additive Schwarz\\Block Jacobi} & BoomerAMG & \makecell[c]{BFBT\\Additive Schwarz\\BoomerAMG} & Custom \cite{esmaily2013new,esmaily2015bi} & \makecell{FAC\\Block Jacobi\\Block Gauss-Seidel} & Jacobi \\\hline
        \makecell[r]{\textbf{Backends and}\\\textbf{dependencies}} & \makecell{deal.II, Trilinos,\\PETSc, Boost,\\VTK} & \makecell{MKL,\\SuperLU,\\Hypre} & Hypre & --- & PETSc & FEniCS, PETSc
    \end{tabular}
    \caption{Comparison of popular CFD software for cardiovascular simulations.}
    \label{tab:cfd_software_comparison}
\end{table}
\end{landscape}

A more detailed comparison about the modeling and numerical features of the previously mentioned solvers is presented in \cref{tab:cfd_software_comparison}. Since ExaDG, Lethe, OpenFOAM, and FEniCS are designed as general-purpose and highly-customizable FEM libraries, they have not been included in the comparison table.

Compared to these alternatives, our \lifexcfd{} solver offers several distinctive advantages, mostly inherited from the \lifex{} core structure \cite{africa2022flexible}. It is designed to be user-friendly and easy to use, even for biomedical researchers without extensive experience in computational fluid dynamics, and comes with a clean and meticulously documented code base. 
The solver is implemented in C++ using modern programming paradigms and leverages MPI for distributed memory parallelism. Moreover, it supports arbitrary finite elements (among those available in \dealii{}), the possibility to import arbitrary meshes with either hexahedral or tetrahedral elements and several linear algebra backends (namely Trilinos \cite{trilinos-website}, PETSc \cite{petsc-user-ref} and \dealii{} itself), ensuring a fine control over the numerical setting. The solver is also designed to be scalable, allowing it to efficiently simulate large-scale scenarios. In addition to the numerical and programming features stemming from its foundation on \lifex{}, \lifexcfd{} can easily and efficiently solve differential problems with moving geometries, accounting for both moving domain boundaries, such as vessel walls, and immersed surfaces like valves. This unique capability sets \lifexcfd{} apart from other CFD packages.
Furthermore, it supports several types of boundary conditions, all of which can be easily accessed and configured through the external parameter files and used without the need to access and modify the source code.
To the best of our knowledge, none of the packages mentioned above exhibits similar features all together.

Besides the aforementioned programming features and numerical methods implemented, a paramount value proposition of \lifexcfd{} is the seamless integration with several other solvers for the cardiac function based on \lifex{} \cite{africa2023lifex_fiber}, such as electrophysiology \cite{africa2023matrix}, mechanics \cite{cicci2022deep, cicci2022efficient, cicci2022projection}, electromechanics \cite{fedele2023comprehensive, piersanti20223d, regazzoni2022cardiac, salvador2021electromechanical}, CFD \cite{corti2022impact, zingaro2022geometric, zingaro2023electromechanics, fumagalli2022image, marcinno2022computational, bennati2023image, bennati2023turbulence, zingaro2022modeling}, fluid-structure interaction \cite{bucelli2022partitioned}, electro-mechano-fluid interaction \cite{bucelli2022mathematical, bucelli2022stable} and myocardial perfusion \cite{di2022prediction, zingaro2023comprehensive}, which have been exploited ranging from single-chamber to whole-heart simulations.

The paper is organized as follows. In Sec.~\ref{sec:models-methods}, we describe the underlying mathematical formulation and the corresponding numerical algorithms used in our solver, including its ability to handle complex geometries, enforce several kinds of boundary conditions, and simulate both laminar and turbulent flows with efficient linear solvers and preconditioners. Implementation details and instructions on how to run \lifexcfd{} simulations are described in Sec.~\ref{sec:implementation}. We verify our implementation by running convergence analyses and by assessing its parallel performance on the \textit{Beltrami flow} benchmark \cite{ethier1994exact}; we also demonstrate the code accuracy and reliability in reproducing different physiological flow regimes through a series of tests inspired by patient-specific vascular and cardiac applications of increasing complexity. All the numerical results are reported in Sec.~\ref{sec:numerical-results}. Finally, some conclusive remarks and the impact of \lifexcfd{} are discussed in Sec.~\ref{sec:conclusions}.

\section{Mathematical and numerical methods}
\label{sec:models-methods}

In the following sections, we introduce the mathematical models and numerical methods characterizing the \lifexcfd{} solver.

\subsection{Mathematical models}
\label{sec:models}

In large vessels and heart chambers, blood can be regarded as a Newtonian incompressible fluid, in spite of the presence of small particles suspended and carried by the plasma. Therefore, we model it with the \textit{incompressible Navier-Stokes equations} \cite{perktold1994flow, quarteroni2017cardiovascular, taylor1996finite, taylor1998finite}, using the ALE formulation \cite{donea1982arbitrary, donea2004arbitrary, hughes1981lagrangian} to account for the motion of vessels and heart chambers.
Moving immersed valves are modeled by the Resistive Immersed Implicit Surface (RIIS) method \cite{fedele2017patient,fumagalli2020image}.

\subsubsection{Domain displacement}\label{sec:lifting}

Let $(0, T)$ be the temporal domain, with $T$ denoting the final time. Let $\Omega_t \subset \mathbb{R}^3$ be the spatial fluid domain at time $t \in (0, T)$, let $\Gamma_t = \partial \Omega_t$ be its boundary, and let $\bm n$ be the outward-directed normal unit vector. To model the displacement of the domain over time, we introduce a fixed reference configuration $\widehat\Omega \subset \mathbb{R}^3$, such that the domain in current configuration $\Omega_t$ is defined at any $t \in (0, T)$ as
\begin{equation*}
    \Omega_t  = \{ \bm x \in \mathbb R^3: \; \bm x = \mathcal A_t(\xhat) = \xhat + \displ(\xhat, t), \; \xhat  \in \widehat \Omega \},
\end{equation*}
where $\displ: \widehat{\Omega} \times (0, T) \to \mathbb R^3$ is the domain displacement. Throughout this paper, we assume that the displacement $\displ_{\widehat{\Gamma}^\mathrm{D}}: {\widehat{\Gamma}^\mathrm{D}} \times (0, T) \to \mathbb R^3$ is known on a Dirichlet boundary portion of $\partial \widehat \Omega$, that we denote by $\widehat{\Gamma}^\mathrm{D}$. We recover the displacement in the bulk domain by solving the following lifting problem:
\begin{equation} \begin{cases}
    -\div(\alpha\, \sigma^\mathrm{L}(\displ)) = \bm 0 &  \text{ in } \widehat \Omega \times (0, T), \\
	\displ = \displ_{\widehat{\Gamma}^\mathrm{D}} &  \text{ on }  { \widehat \Gamma^\mathrm{D}} \times (0, T), \\
	\sigma^\mathrm{L}(\displ) \widehat{\bm n} = \bm 0 & \text{ on }  \partial \widehat \Omega \setminus \widehat{\Gamma}^\mathrm{D} \times (0, T).
\end{cases} \label{eq:lifting} \end{equation}
Concerning the choice of $\sigma^\mathrm{L}(\displ)$, \lifexcfd{} provides the following options:
\begin{itemize}
    \item \textit{harmonic extension}: $\sigma^L (\displ) = \grad\displ$;
    
    \item \textit{linear elasticity}: 
    \[ \sigma^L(\displ) = \frac{E}{1 + \nu} \grad^S \displ + \frac{E \nu}{(1+\nu)(1-2\nu)} \left(\div \displ\right) I\;, \]
    where $E$ and $\nu$ are the Young and Poisson moduli of a fictitious linear elastic material, and $\grad^S$ denotes the symmetric gradient, defined as 
    \[ \grad^S\displ = \frac{1}{2}\left(\grad\displ+\grad\displ^T\right)\;. \]
\end{itemize}

The coefficient $\alpha: \widehat{\Omega} \to \mathbb R$ in \eqref{eq:lifting} is introduced to locally increase the stiffness in critical areas of the domain, to preserve the quality of mesh elements. \lifexcfd{} offers three options:
\begin{itemize}
    \item no stiffening, i.e. $\alpha = 1$ in $\widehat{\Omega}$;
    
    \item \textit{Jacobian-based stiffening} \cite{stein2003mesh}:
    \begin{equation*}
        \alpha(\xhat) = \left(\frac{J_0}{J(\xhat)}\right)^\chi\;,
    \end{equation*}
    where $J_0$ and $\chi$ are positive user-defined parameters, and $J(\xhat)$ is the Jacobian of the map from the unit reference element to the one in $\widehat\Omega$; with this approach, elements that are small are made stiffer;
    
    \item \textit{boundary-based stiffening} \cite{jasak2006automatic}:
    \begin{equation*}
        \alpha(\xhat) = \max(d(\xhat), \; \beta)^{-\gamma}\;,
    \end{equation*}
    where $d(\xhat)$ is the distance from the boundary (or a portion of the boundary), and $\beta$ and $\gamma$ are two user-defined positive parameters; with this approach, the stiffness is increased in the elements that are close to the moving boundary.
\end{itemize}

Finally, the ALE velocity is computed by taking the derivative of the displacement w.r.t.~time:
\begin{equation}
	\uale = \pdv{{\displ}}{t}\circ\mathcal{A}_t^{-1}\ \; \; \mathrm{in} \; \Omega_t \times (0, T). 
	\label{eq:velocity-ALE}
\end{equation}

\subsubsection{Incompressible Navier-Stokes equations}

We model blood as a fluid of constant density $\rho$ and constant dynamic viscosity $\mu$.
Let $\vel:\, \Omega_t \times (0, T) \to \mathbb R^3$ be the fluid's velocity and $p:\, \Omega \times (0, T) \to \mathbb R$ be its pressure. Their evolution is prescribed by the incompressible Navier-Stokes equations in the ALE framework with RIIS modeling of immersed structures such as cardiac valves, that is:
\begin{subnumcases}{\label{eq:ns-ale-riis}}
    \begin{multlined} \rho \frac{\hat\partial \vel}{\hat\partial t} + \rho((\vel - \uale)\cdot \grad )\vel + \div \sigma (\vel, p) \\[-0.5em] + \resistive = \bm f \end{multlined} & in $\Omega_t \times (0, T)$, \label{eq:ns-ale-riis-momentum} \\[0.5em]
    \div \vel = 0 & in $\Omega_t \times (0, T)$, \label{eq:ns-ale-riis-continuity} \\[0.5em]
    \vel = \vel_0 & in $\Omega_0 \times \{0\}$. \label{eq:ns-ale-riis-initial-condition}
\end{subnumcases}
In the above, $\frac{\hat\partial}{\hat\partial t}$ denotes the time derivative in the ALE framework \cite{donea2004arbitrary}, $\bm f:\Omega_t \times (0, T) \to \mathbb{R}^3$ is a forcing term, and $\vel_0:\Omega_0 \to \mathbb{R}^3$ is a suitable initial condition. The latter can either be zero, or be imported from clinical data or other simulation results. The tensor $\sigma$ is the Cauchy stress tensor, defined for an incompressible, viscous and Newtonian fluid as 
\begin{equation*}
\sigma (\vel, p) = -p I + 2\mu \grad^S \vel.
\end{equation*}
The alternative definition $\sigma(\vel,p)=-p I+\mu\grad\vel$ can also be considered: see \eqref{eq:diffusion-term} in \cref{sec:methods}.
System \eqref{eq:ns-ale-riis} must be endowed with suitable boundary conditions, as described in \cref{sec:bcs}.

The term $\resistive$ in \eqref{eq:ns-ale-riis-momentum} is introduced to account for the presence of heart valves in the fluid in cardiac simulations by means of the \textit{Resistive Immersed Implicit Surface} (RIIS) method  \cite{fedele2017patient, fumagalli2020image}.
The valves are represented by a set $\mathcal{I}_{\mathrm v}$ of immersed surfaces $\sigmak, \mathrm{k}\in\mathcal{I}_{\mathrm v}$. Each of them is characterized by a resistance coefficient $R_\k$ and a parameter $\varepsilon_\k$ representing the half thickness of the valve leaflets. To each valve we associate a signed distance function $\varphi_\k: \Omega_t \times (0, T) \to \mathbb R$. Then, $\resistive$ is defined as \cite{ fedele2017patient, fumagalli2020image, bucelli2022mathematical, zingaro2022geometric}
\begin{equation}\label{eq:riis}
    \resistive= \sum_{\k \in \mathcal{I}_{\mathrm v}} \frac{R_\mathrm{k}}{\varepsilon_\mathrm{k}} \delta_{\sigmak, \varepsilon_\mathrm{k}}(\varphi_\mathrm{k})\left(\vel - \uale - \vel_{\sigmak}\right ).
\end{equation}
This term weakly imposes the no-slip condition on the valve surfaces, by penalizing the difference between the relative fluid velocity $\vel - \uale$ and the relative velocity of the valves' leaflets $\vel_{\sigmak}$, which is prescribed. The term $\bm {\mathcal R}$ has support over a narrow layer around $\sigmak$ defined through the following smoothed Dirac delta function:
\begin{equation*}
    \delta_{\sigmak, \varepsilon_\k} (\varphi_\k(\bm x)) = 
    \begin{cases}
    \dfrac{1+\cos(\pi \varphi_\k(\bm x)/\varepsilon_\k))}{2\varepsilon_\k} & \text{if } |\varphi_\k(\bm x)| \leq \varepsilon_\k,
    \\
    0 & \text{if } |\varphi_\k(\bm x)| > \varepsilon_\k,
    \end{cases}
\end{equation*}
for all $\k \in\mathcal I_{\mathrm v}$.

\subsection{Boundary conditions}
\label{sec:bcs}

\lifexcfd{} supports different types of boundary conditions, and for each type several options are available. In the following sections, we briefly describe each of the implemented options. All the boundary condition types discussed below can be combined seamlessly, by decomposing the boundary in an arbitrary number of subsets and selecting one of the following options for each subset. All the coefficients and parameters for the options that follow are easily customizable by the user without modifying the source code.

\subsubsection{Dirichlet boundary conditions}
\label{sec:bcs-dirichlet}

Let $\gammad \subset \partial\Omega_t$ denote a subset of the domain boundary.\lifexcfd{} supports no-slip boundary conditions, i.e. 
\[ \vel = \uale \qquad \text{on } \gammad \times (0, T)\;, \]
or $\vel = \bm 0$ if the domain is not moving (i.e. if the ALE formulation is not enabled).
Alternatively, it is possible to prescribe Dirichlet conditions in the form
\begin{equation*}
    \vel(\mathbf x, t) = \bm g(\mathbf x, t) = \theta g(t)\bm s(\bm x) \qquad \text{on } \gammad \times (0, T)\;,
\end{equation*}
where $\bm g : \gammad \times (0, T) \to \mathbb{R}^3$ is expressed by separation of variables, with $g:(0, T) \to \mathbb R$ and $\bm s: \gammad \to \mathbb R^3$. The parameter $\theta>0$ is a \textit{repartition factor} that can be used to split total flow rate over multiple boundaries.

\lifexcfd{} offers several options for the function $g(t)$:
\begin{itemize}
    \item A \textit{constant function}: 
    \begin{equation*}
        g(t) = \overline{g}, \text{ for all } t \in (0, T).
    \end{equation*}
    
    \item A sinusoidal \textit{pulsatile function} from $g_\mathrm{min}$ to $g_\mathrm{max}$, with period $\mathcal T$:
    \begin{equation*}
    g(t) = g_{\min} + (g_{\max} - g_{\min})\;\frac{1}{2} \left (  1 - \cos(\frac{2 \pi  t}{\mathcal T}) \right ). 
    \end{equation*}
    
    \item A \textit{sinusoidal ramp} from $g_\mathrm{0}$ to $g_\mathrm{1}$, starting at $t_\mathrm{0}>0$ and lasting $t_\mathrm{L}>0$:
    \begin{equation*}
    g(t) = 
        \begin{cases}
    g_\mathrm{0} & \text{ if } t < t_\mathrm{0},
    \\
    g_\mathrm{0} + \frac{g_\mathrm{1} - g_\mathrm{0}}{2} \left ( 1 - \cos \left ( \frac{\pi (t - t_\mathrm{0})}{t_\mathrm{L}}\right )  \right ) & \text{ if } t_\mathrm{0} \leq t < t_\mathrm{0} + t_\mathrm{L},
    \\
    g_\mathrm{1} & \text{ if } t \geq t_\mathrm{0} + t_\mathrm{L}.
    \end{cases}
    \end{equation*}
    
    \item \textit{Interpolation in time} of prescribed flow rate or velocity data: the function $g(t)$ is obtained by interpolating a given a set of pairs $\{(t_i, g_i)\}_{i = 0}^{N_g}$. The interpolation can be a piecewise linear polynomial, a piecewise cubic spline, or a Fourier series interpolation \cite{quarteroni2010numerical}.
\end{itemize}

\bigskip

For the space-dependent function $\bm s(\bm x)$, the options below are available.
\begin{itemize}
    \item A \textit{parabolic profile} for a circular boundary of radius $\widetilde{R}$:
    \begin{equation*}
        \bm s(\bm x) = - 2 \frac{\widetilde{R}^2 - r^2(\bm x)}{\pi \widetilde{R}^4} \bm n(\bm x),
    \end{equation*}
    where $r(\bm x)$ is the distance of $\bm x$ from the barycenter of $\gammad$. A parabolic profile can be used for instance to prescribe a laminar inlet velocity profile \cite{kundu2015fluid}.
    
    \item A \textit{uniform profile}, i.e. a prescribed inflow velocity that is normal to the $\gammad$ and constant in space: 
    \begin{equation*}
        \bm s(\bm x) = - \bm n (\bm x).
    \end{equation*}
    A uniform profile can be used, for instance, to prescribe a turbulent inlet velocity profile \cite{kundu2015fluid}.
    
    \item A \textit{Womersley profile} \cite{womersley1955method} for a circular boundary of radius $\widetilde{R}$, commonly used to describe a pulsatile flow in the cardiovascular system \cite{quarteroni2017cardiovascular, womersley1955method}. This profile is computed by solving the inverse problem\footnote{This inverse problem is solved once and for all at the beginning of the simulation.} described in \cite{berselli2013exact}.
    The Womersley profile is only compatible with $g(t)$ defined by interpolating given data using a Fourier series.
\end{itemize}

\begin{remark}
Parabolic and Womersley profiles have unit flow rate, so that the function $g(t)$ represents flow rate over time. Conversely, the uniform profile has unit velocity, and the function $g(t)$ is a velocity over time.
\end{remark}

\subsubsection{Periodic boundary conditions}
\label{sec:bcs-periodic}

Periodic boundary conditions are also available. Let $\Gamma^\mathrm{A} \subset \partial\Omega$ and $\Gamma^\mathrm{B} \subset \partial\Omega$ be two planar subsets of the boundary of the domain, mutually mapped through a translation $\bm\phi_\mathrm{AB}:\mathbb{R}^3 \to \mathbb{R}^3$ such that $\Gamma^\mathrm{B} = \bm\phi_\mathrm{AB}(\Gamma^\mathrm{A})$. Then, periodic conditions are expressed as
\begin{equation*}
    \vel(\bm x, t) = \vel(\bm\phi_\mathrm{AB}(\bm x), t) \qquad \text{on } \Gamma^\mathrm{A}\times(0, T)\;.
\end{equation*}

Periodic conditions are only supported for hexahedral meshes. Moreover, the discretizations of $\Gamma^\mathrm{A}$ and $\Gamma^\mathrm{B}$ are required to be conforming (i.e. there must be a one-to-one correspondence between mesh elements on the two surfaces).

\subsubsection{Neumann boundary conditions}\label{sec:bcs-neumann}
\lifexcfd{} supports Neumann boundary conditions in the form
\begin{equation}
    \sigma(\vel, p)\bm n = -h(t) \bm n \qquad \text{on } \gamman\times(0,T)\;,
    \label{eq:bcs-neumann}
\end{equation}
where $\gamman \subset \partial\Omega_t$, $\bm n$ is the outgoing normal to $\gamman$ (at a fixed time) and $h(t): (0, T) \to \mathbb R$ is a time-dependent function representing the space-averaged pressure on the Neumann boundary. The same options introduced for $g(t)$ (\cref{sec:bcs-dirichlet}) are available for $h(t)$.

Optionally, backflow instabilities can be prevented on Neumann boundaries by enabling \textit{inertial backflow stabilization} \cite{bertoglio2014tangential, moghadam2011comparison, vignon2010outflow}, i.e. by modifying \eqref{eq:bcs-neumann} as
\begin{equation*}
    \sigma(\vel, p)\bm n = -h(t) \bm n + \beta\frac{\rho}{2}|\bm u\cdot\bm n|_{-}\vel \qquad \text{on } \gamman\times(0,T)\;,
\end{equation*}
where $|\vel\cdot\bm n|_{-} = \min\{\bm u\cdot\bm n, 0\}$ and $\beta > 0$ is a user-defined stabilization parameter \cite{bertoglio2014tangential, moghadam2011comparison, vignon2010outflow}.

\subsubsection{Resistance boundary conditions}
\label{sec:bcs-resistance}

Resistance boundary conditions \cite{bazilevs2009patient, bazilevs2008isogeometric,quarteroni2016geometric,dede2021computational} are conditions of the form
\begin{equation*}
    \sigma(\vel, p)\bm n = -\left(p_0(t) + C_\mathrm{R}\int_{\gammar}(\vel - \uale)\cdot\bm n\,d\gamma \right) \bm n \qquad \text{on } \gammar\times(0,T)\;,
\end{equation*}
where $\gamma \subset \partial\Omega$, $p_0(t)$ is a baseline pressure, $C_\mathrm{R}>0$ the resistance coefficient. The term $p_0(t)$ can be defined using the same temporal functions introduced for $g(t)$.

\subsubsection{Free-slip boundary conditions}
Free-slip conditions, also known as symmetry conditions, are mixed Dirichlet-Neumann conditions of the form
\begin{align*}
    \vel\cdot\bm n = 0 & \qquad \text{on } \gammad \times (0, T)\;, \\
    \left(\sigma(\vel, p)\bm n\right)\cdot \bm t_i = 0 & \qquad \text{on } \gammad \times(0, T) \qquad i \in \{1, 2\}\;,
\end{align*}
where $\bm t_i$, for $i \in \{1, 2\}$ are the tangent unit vectors to $\gammad$.

\subsection{Numerical methods}
\label{sec:methods}
We introduce the infinite-dimensional function spaces 
\begin{equation*}
	\spacevelocity := \, \left\{ \bm v \in [H^1(\Omega_t)]^3: \bm v = \bm g \text{ on } \gammad\right\}, \quad 
	\spacepressure:= \,  L^2(\Omega_t).
	\label{eq:function_spaces_ale_riis}
\end{equation*}

For the spatial discretization of \eqref{eq:ns-ale-riis}, we introduce a mesh $\mathcal T^h$ over $\Omega_t$, composed of either tetrahedra or hexahedra (\lifex{} supports both types of meshes). We introduce the function space of \textit{Finite Elements} (FE) with piecewise Lagrangian polynomials of degree $r\geq 1$ over $\mathcal T^h$ as 
\begin{equation*}
	X_r^h = \{ v_h \in C^0(\overline{ \Omega}):\,  v_h|_K \in \mathbb{P}^r,\,  \text{ for all } K \in \mathcal T^h\},
\end{equation*}
wherein $h$ is the diameter of the grid element $K \in \mathcal T^h$. We introduce the finite-dimensional spaces $\spacevelocityh = \spacevelocity \cap [X_r^h]^3$ and $ \spacepressureh = \spacepressure \cap X_r^h$, and denote by $ \velh \in \spacevelocityh$, $p_h \in \spacepressureh$ the FE approximations of $\vel$ and $p$, respectively. \lifexcfd{} supports polynomials of order \num{1} and \num{2} on tetrahedral meshes, and polynomials of arbitrary degree on hexahedral meshes.

Temporal discretization is carried out through \textit{Backward Differentiation Formulas} (BDF) \cite{quarteroni2010numerical, curtiss1952integration} of order $\sigma_\bdf = 1, 2, 3$. We partition the time domain $(0, T)$ into $N_t$ subintervals of equal size $\dt = T/{N_t}$, and we denote with the superscript $n$ quantities related to the time-step $n$, with $n=0, \dots, N_t$. The approximation of the velocity time derivative reads \cite{quarteroni2010numerical, forti2015semi}:
\begin{equation}
	\pdv{\velh}{t} \bigg \vert_{t^{n+1}} \approx \frac{\alpha_\bdf \velhcurrent - \velhbdf}{\dt} \quad \text{for } n = \sigma_\bdf - 1, \dots, N_t - 1. 
	\label{eq:bdf}
\end{equation}
wherein $\alpha_\bdf$ is a coefficient depending on the order $\sigma_\bdf$ of the method and $\velhbdf$ is a linear combination of velocities at timesteps $n$, $n-1$, $\dots$, $n - \sigma_\bdf + 1$.

We also introduce an extrapolated velocity $\velhext$ that approximates $\velhcurrent$ with accuracy order $\sigma_\bdf$ through a linear combination of velocities at timesteps $n$, $n-1$, $\dots$, $n - \sigma_\bdf + 1$. We refer to \cite{forti2015semi} for the definition of $\alpha_\bdf$, $\velhbdf$ and $\velhext$.

The advection term in \eqref{eq:ns-ale-riis-momentum} can be formulated in a \textit{fully-implicit} or \textit{semi-implicit} \cite{forti2015semi} way. For the sake of a compact notation, we denote the advection velocity by $\ustar$, defined as
\begin{equation*}
	\ustar = \begin{cases}
		\velhcurrent - \ualehcurrent  & \text{ (implicit formulation)}, \\
		\velhext- \ualehcurrent & \text{ (semi-implicit formulation)}. \\
	\end{cases}
	\label{eq:u_star}
\end{equation*}
The time-discrete mesh velocity $\ualehcurrent \approx \pdv{\displ}{t}$ is evaluated using the BDF method \eqref{eq:bdf}, of the same order used to approximate the time derivative of $\velh$.

The fully-discrete formulation of \eqref{eq:ns-ale-riis} reads:
given $\velh^n, \dots, \velh^{n+1-\sigma_\bdf}$, for any $n = 0$, $\dots$, $N_t -1$, find $(\velhcurrent, \phcurrent)  \in \spacevelocityh \times \spacepressureh $ such that, for all $(\vh, \qh) \in \spacemomentumtesth \times \spacepressureh$,
\begin{equation}
	\begin{aligned}
		& \left ( \rho \frac{\alpha_\bdf \velhcurrent}{\dt} , \vh \right )_{\Omega_{n+1}} + 
		\left ( \rho \left ( \ustar  \cdot \grad \right ) \velhcurrent , \vh \right)_{\Omega_{n+1}} + 
            \mathcal D(\velhcurrent, \vh )
            \\ 
        & \quad - \left ( \phcurrent , \div \vh \right)_{\Omega_{n+1}} + 
		\left ( \div \velhcurrent , \qh \right )_{\Omega_{n+1}}  
		+ \left ( \bm{\mathcal R}(\velhcurrent, \ualehcurrent ), \vh \right )_{\Omega_{n+1}}  
		\\
		& \quad + \sum_{K\in\mathcal T_h}\mathcal{S}^K(\velhcurrent, \ustar, \phcurrent, \phext, \vh, \qh)_{\Omega_{n+1}}
            \\
            & \quad =  \left ( \bm f^{n+1}, \vh \right ) +
		\left ( \bm h^{n+1} , \vh \right )_{\Gamma^\mathrm{N}_{n+1}} +  \left ( \rho \frac{\velhbdf}{\dt}, \vh \right )_{\Omega_{n+1}}\;.
	\end{aligned}
	\label{eq:fullydiscrete_ns_ale_riis_vmsles_quasistatic_unified}
\end{equation}
In the above, $(\cdot, \cdot)_{X}$ denotes the scalar product in $L^2(X)$.
$\mathcal D$ is the diffusion term, for which \lifexcfd{} offers the following formulations:
\begin{equation}
    \mathcal D (\velhcurrent, \vh) = 
    \begin{cases}
        \left ( \mu \grad \velhcurrent, \grad \vh \right )_{\Omega_{n+1}} & \text{ (\textit{gradient - gradient})} \\
        \left ( 2 \mu \grad^S \velhcurrent, \grad \vh \right )_{\Omega_{n+1}} & \text{ (\textit{symmetric gradient - gradient})} \\
        \left ( 2 \mu \grad^S \velhcurrent, \grad^S \vh \right )_{\Omega_{n+1}} & \text{ (\textit{symmetric gradient - symmetric gradient})}
    \end{cases}
    \label{eq:diffusion-term}
\end{equation}
The formulation \eqref{eq:fullydiscrete_ns_ale_riis_vmsles_quasistatic_unified} includes terms $\mathcal S^K$ that can be optionally enabled to stabilize the Galerkin formulation. \lifexcfd{} provides the \textit{Streamline Upwind Petrov Galerkin - Pressure Stabilizing Petrov Galerkin} (SUPG-PSPG) \cite{tezduyar2003stabilization} and the \textit{Variational Multiscale - Large Eddy Simulation} (VMS-LES) \cite{forti2015semi, bazilevs2007variational, takizawa2014st} stabilization methods. Both methods stabilize the problem in terms of the \textit{inf-sup} condition and stabilize possible advection-dominated regimes (typically occurring in cardiovascular hemodynamics) \cite{quarteroni2017numerical}. Furthermore, the VMS-LES acts also as a turbulence model, allowing to model the transition-to-turbulence regime characterizing the cardiovascular blood flow \cite{chnafa2014image, nicoud2018large, zingaro2021hemodynamics}. In a compact form, we define the stabilization terms as:
\begin{equation}
    \mathcal{S}^K(\velhcurrent, \ustar, \phcurrent, \phext, \vh, \qh) = 
    \begin{cases}
        \mathcal{S}^K_\mathrm{SUPG-PSPG}(\velhcurrent, \ustar, \phcurrent, \vh, \qh) & \text{ if SUPG-PSPG} \\[1.0em]
        \begin{aligned}
            & \mathcal{S}^K_\mathrm{SUPG-PSPG}(\velhcurrent, \ustar, \phcurrent, \vh, \qh) \\
            & \;\; + \mathcal{S}^K_\mathrm{VMS}(\velhcurrent, \ustar, \phcurrent, \vh) \\
            & \;\; + \mathcal{S}^K_\mathrm{LES}(\velhcurrent, \ustar, \phcurrent, \phext, \vh)
        \end{aligned} & \text{ if VMS-LES}
    \end{cases}
    \label{eq:stabilization_form}   
\end{equation}
with the SUPG-PSPG and VMS terms defined, respectively, as
\begin{align}
    \mathcal{S}^K_\mathrm{SUPG-PSPG}(\velhcurrent, \ustar, \phcurrent, \vh, \qh) = & \left ( \, \tau_{\text M}(\ustar) \bm r_{\text M} (\velhcurrent, \phcurrent), \rho \ustar  \cdot \grad \vh + \grad \qh\right )_K & \notag \\
    & +  \left (\tau_{\text C}(\ustar)r_{\text C}(\velhcurrent), \div \vh\right)_K\;,
    \label{eq:stabilization_form_SUPG} 
    \\
    \mathcal{S}^K_\mathrm{VMS}(\velhcurrent, \ustar, \phcurrent, \vh) = & \left( \tau_{\text M}(\ustar)\bm r_{\text M}(\velhcurrent, \phcurrent), \rho \ustar \cdot (\grad \vh)^T \right )_K\;.
    \label{eq:stabilization_form_VMS} 
\end{align}
The LES term -- modeling the Reynolds stresses -- is defined according to the implicit or semi-implicit formulation, as \cite{forti2015semi, bazilevs2007variational}:
\begin{align}
    & \mathcal{S}^K_\mathrm{LES}(\velhcurrent, \ustar, \phcurrent, \phext, \vh) = \notag &&\\
    & \begin{cases}
    - ( \tau_{\text M}(\ustar)\bm r_{\text M}(\velhcurrent, \phcurrent) \otimes 
    \tau_{\text M}(\ustar)\bm r_{\text M}(\velhcurrent, \phcurrent) , \rho \grad \vh)_K  & \text{(implicit formulation),} 
    \\
    - ( \tau_{\text M}(\ustar)\bm r_{\text M}(\velhext, \phext) \otimes 
    \tau_{\text M}(\ustar)\bm r_{\text M}(\velhcurrent, \phcurrent), \rho \grad \vh )_K   & \text{(semi-implicit formulation).}  
    \end{cases} \label{eq:stabilization_form_LES} &&
\end{align} 
In the above, $\bm r_\mathrm{M}$ and $r_\mathrm{C}$ are the strong residuals of the momentum balance \eqref{eq:ns-ale-riis-momentum} and continuity \eqref{eq:ns-ale-riis-continuity} equations, defined on each mesh element $K$ as
\begin{subequations}
\begin{align*}
\bm r_\text{M}^K( \velhcurrent, \phcurrent)  =  &
\rho \left (\frac{\alpha_\bdf \velhcurrent - \velhbdf}{\dt} + 
\ustar \cdot \grad \velhcurrent \right )+ \grad \phcurrent  - \mu \laplacian \velhcurrent  
\\ & 
+  \bm{\mathcal R}(\velhcurrent, \ualehcurrent) - \bm f^{n+1},
\\ 
r_\text{C}^K(\velhcurrent) = & \div \velhcurrent.
\end{align*}
\end{subequations}
The stabilization parameters are defined as \cite{zingaro2022geometric, forti2015semi}
	\begin{align*}
	\tau_\text{M} (\ustar) & = 
	\left ( 
	\frac{\sigma_\bdf^2\rho^2}{\dt^2} + 
	\rho^2 \ustar \cdot \overline{G} \ustar +
	C_r \mu^2  \overline{G} : \overline{G} + \sum_{\mathrm{k}\in \mathcal I_\mathrm{v}}\frac{R_k^2}{\varepsilon_\mathrm{k}^2}\delta^2_{\Sigma_\mathrm{k}, \varepsilon_\mathrm{k}}(\varphi_\mathrm{k})
	\right)^{-\frac{1}{2}},
	\\
	\tau_\text{C} (\ustar) & = 
	\left(\tau_\text{M} (\ustar) \overline{\bm g} \cdot \overline{\bm g}\right)^{-1}\;,
	\end{align*}
where $\overline{G}=J^{-T} J^{-1}$ and $\overline{\bm g} = J ^{-T} \bm 1.$ are the metric tensor and the metric vector, respectively, {depending on the Jacobian $J$ of the map from the reference element to the current one \cite{bazilevs2007variational}}. We remark that the definition of the stabilization terms has been modified with respect to the original formulation of \cite{bazilevs2007variational} by accounting for the presence of immersed surfaces \cite{zingaro2022geometric}.

An analogous FE approximation is introduced for the quasi-static lifting problem \eqref{eq:lifting},\linebreak through which both the discrete displacement $\displ_h$ and the discrete ALE velocity $\uale_h$ are defined.

\subsection{Linear algebra}\label{sec:linearalgebra}

Following space and time discretizations, the lifting problem \eqref{eq:lifting} leads to a symmetric, positive definite linear system, which we solve using either the GMRES or the conjugate gradient (CG) method \cite{quarteroni2010numerical, saad2003iterative}. The system can be preconditioned using algebraic multigrid (AMG) \cite{xu2017algebraic}, additive Schwarz \cite{quarteroni1999domain} or a block Jacobi preconditioner.

The discretization of Navier-Stokes equations \eqref{eq:fullydiscrete_ns_ale_riis_vmsles_quasistatic_unified} leads to a block algebraic system, which can be linear or non-linear depending on the implicit or semi-implicit treatment of the advection term. It is linearized through Newton's method \cite{quarteroni2017numerical}\footnote{In the linear case, only one Newton iteration is performed.}, and the resulting block linear system is solved through the GMRES method \cite{quarteroni2010numerical, saad2003iterative}, using the aSIMPLE preconditioner \cite{deparis2014parallel}. For further details on the algebraic form of the problem and its block structure, see \cite{deparis2014parallel}. The approximation of the velocity and Schur complement blocks is obtained through AMG, additive Schwarz or block Jacobi \cite{saad2003iterative}.

\section{Implementation}
\label{sec:implementation}

\lifexcfd{} builds upon the core functionalities of \lifex{} \cite{africa2022flexible}, which in turn relies on \dealii{} \cite{dealii2019design, dealII93} to provide general-purpose software facilities for finite elements. Specifically, \lifexcfd{}
implements the mathematical models, numerical methods and utilities needed for CFD simulations in the cardiovascular setting. The linear algebra backend is offered either by Trilinos \cite{trilinos-website} or by PETSc \cite{petsc-web-page,petsc-user-ref}, as wrapped by \dealii{}. This includes the implementation of the linear solvers (CG and GMRES) and of the black-box preconditioners (AMG, additive Schwarz, block Jacobi) that \lifex{} supports. Other dependencies include VTK \cite{schroeder2006visualization}, the Boost libraries \cite{schaling2011boost} and general-purpose compilation tools such as CMake\footnote{\url{https://cmake.org/}} and GNU make. We provide a Docker\footnote{\url{https://www.docker.com/}} image bundling all the dependencies, to facilitate the installation of \lifexcfd{} on new systems. We refer to the \href{\documentationurl}{online documentation}\footnote{\url{\documentationurl}} for further instructions on the download and installation procedures.

\lifex{} can read and write hexahedral and tetrahedral meshes in the widely used \texttt{msh} format. This format is supported by many mesh generation software, such as \texttt{Gmsh}\footnote{\url{http://gmsh.info}}, \texttt{Netgen}\footnote{\url{https://ngsolve.org/}}, \texttt{vmtk}\footnote{\url{http://www.vmtk.org}} and \texttt{meshtool}\footnote{\url{https://bitbucket.org/aneic/meshtool/src/master/}}.
Other mesh formats can be converted to \texttt{msh} using, for instance, the open-source library \texttt{meshio}\footnote{\url{https://github.com/nschloe/meshio}}.

We exploit forward automatic differentiation \cite{sacado-website} to assemble the linear system associated to \eqref{eq:fullydiscrete_ns_ale_riis_vmsles_quasistatic_unified}. This is especially useful to deal with the large number of terms associated to the SUPG-PSPG and VMS-LES stabilizations \eqref{eq:stabilization_form}. In particular, this allows to easily implement the implicit VMS-LES formulation, in spite of the large number of non-linear terms it involves.

We refer the interested reader to \cite{africa2022flexible} for further details on the core functionalities of \lifex{}. Below, we provide a quick-start guide on how to run simulations within \lifexcfd{}. More information can be found in the \href{\documentationurl}{online documentation}.

\subsection{Running simulations in \texorpdfstring{\lifexcfd{}}{lifex-cfd}}
\label{sec:quick-start}

A \textit{ready-to-run} binary package is available for download at \url{https://doi.org/10.5281/zenodo.7852088}, compatible with any recent enough \texttt{x86-64 Linux} distribution, provided that \texttt{glibc} (\url{https://www.gnu.org/software/libc/}) version $\geq$ \texttt{2.28} is installed. The binary file is shipped in \texttt{AppImage} (\url{https://appimage.org/}) format and is named \verb|lifex-cfd-1.5.0-x86_64.AppImage|. For more detailed instructions on how to run the \lifexcfd{} binary, we refer the reader to \cite{africa2023lifex_fiber}.

Experienced users and those aiming at maximum computational efficiency on High Performance Computing (HPC) facilities and cloud platforms are recommended to compile \lifexcfd{} and its dependencies from source. Instructions are available in the \href{\documentationurl}{online documentation}, together with a manual for running simulations. Below we report a quick-start guide to use the solver, and refer the interested reader to the documentation for further information. All the commands hold also for the binary version, by simply replacing \verb|lifex_fluid_dynamics| with \verb|lifex-cfd-1.5.0-x86_64.AppImage|.

After compiling \lifexcfd{}, the main executable can be found within the compilation folder at \linebreak[4]\texttt{apps/fluid\_dynamics/lifex\_fluid\_dynamics}. A brief description of its command line options can be obtained through
\begin{lstlisting}[language=bash]
$ ./lifex_fluid_dynamics -h
\end{lstlisting}

The executable allows to run test cases with an arbitrary number of boundaries. A labeled volumetric mesh of the domain \(\Omega\) must be provided to define a specific partition of the boundary \(\{\Gamma_i\}_{i=1}^{N_b}\) such that \(\cup_{i=1}^{N_b}\Gamma_i \subseteq \partial\Omega\). Each \(\Gamma_i\) corresponds to a distinct section of the boundary and is identified through appropriate surface tags included in the input mesh.

The configuration of the simulation is supplied through a parameter file. The user can generate a template parameter file using the following command:
\begin{lstlisting}[language=bash]
$ ./lifex_fluid_dynamics -g [minimal|full] \
                         [-b <boundary labels>...]
\end{lstlisting}
Here, \texttt{<boundary labels>} is a list of user-defined labels associated with specific boundary conditions to be imposed. The generated parameter file contains a section for each boundary label, each of which includes a \texttt{Tags} parameter. This parameter allows users to define the association of the boundary label to a list of boundary tags from the input mesh, identifying the \(\Gamma_i\) boundaries where that condition should be applied.
In addition to the boundary condition specifications for each element of \texttt{<boundary labels>}, either no-slip or free-slip boundary conditions can be enforced on other boundaries: see \cref{sec:cylinder} for a practical example.

The level of detail of the parameter files can be optionally reduced (by specifying the \texttt{minimal} option) or increased (by specifying the \texttt{full} option). The former is advised for an initial use of \lifexcfd{}, while the latter exposes advanced options such as parameter choices on the stabilization term and turbulence models, additional options for the ALE lifting and the RIIS method, the Jacobian-based or boundary-based stiffening of the lifting, detailed options on linear algebra and preconditioning, among others. If the user does not specify either option, an intermediate verbosity is selected by default. Parameters that are not present in the file retain their default value.
We refer the reader to the \href{\documentationurl}{documentation of the library} for further information.

The parameters are written into a plain text file that provides a list of key-value pairs, grouped in subsections, that describe the configuration for the simulation to be run. Each parameter is supplied with a brief documentation (included in the parameter file itself) that explains its meaning.
Once the user has edited the parameter file, the simulation can be started with:
\begin{lstlisting}[language=bash]
$ ./lifex_fluid_dynamics [-b <boundary labels>...] \
                         [-f parameter_file_name.prm] \
                         [-o output_folder]
\end{lstlisting}
The boundary labels provided when executing must be the same as those used for generating the parameter file. A parallel simulation is started prepending the command with the \texttt{mpirun} or \texttt{mpiexec} wrapper (which may vary depending on the MPI implementation available), e.g.:
\begin{lstlisting}[language=bash]
$ mpirun -n N ./lifex_fluid_dynamics ...
\end{lstlisting}
where \texttt{N} is the desired number of parallel processes. The binary version supports parallel execution via \texttt{MPICH} (\url{https://www.mpich.org/}) version $\geq$ \texttt{4.0}, whereas there is in principle no restriction in the chosen MPI implementation when building \lifexcfd{} from source.

We point out that the parameter file also contains settings that allow to serialize the solution, so that the simulation can be stopped and restarted at a later time. Refer to the \href{\documentationurl}{online documentation} for additional details.

\section{Results and discussion}
\label{sec:numerical-results}

In the following sections, we analyze the convergence and scalability properties of the solver, and propose some application examples to showcase the functionalities of \lifexcfd{}.

For the sake of reproducibility, all parameter files, meshes and auxiliary data associated with the test cases described below can be downloaded from \url{https://doi.org/10.5281/zenodo.7852088}. The following sections report partial listings of the parameter files used in the numerical examples, including only the most significant settings for each test case. We refer the reader to the full parameter files for more details.

\subsection{Software verification and parallel performance (Test I)}
\label{sec:software-verification}

\begin{figure}
    \centering
    \includegraphics[width=0.8\textwidth]{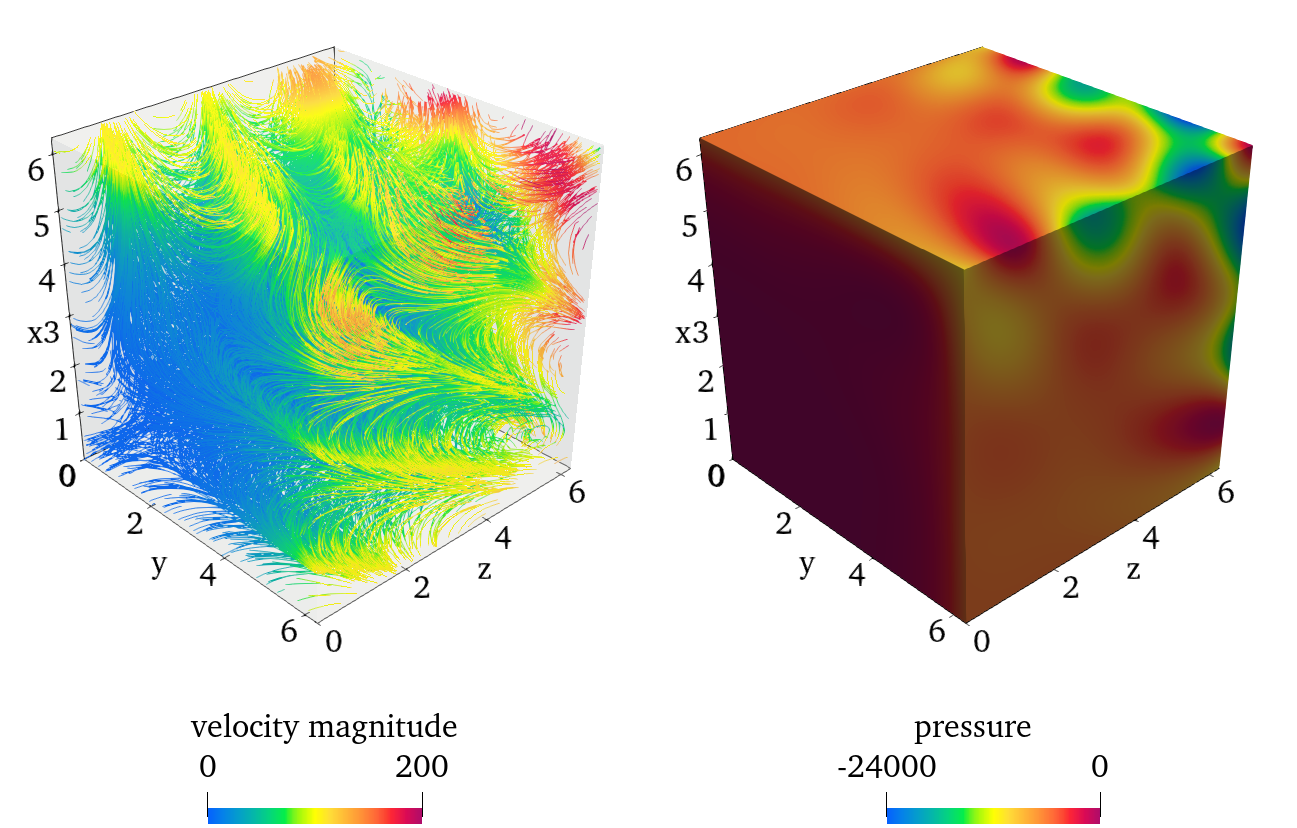}
    \caption{Test I. Solution of the Beltrami flow test case at time $t = 0$. Left: velocity streamlines. Right: pressure.}
    \label{fig:beltrami-solution}
\end{figure}

For the purpose of software verification, we consider a modified version of the \textit{Beltrami flow} benchmark problem \cite{ethier1994exact}, which is a test case on a cubic domain for which an exact analytical solution exists The test case is modified with respect to \cite{ethier1994exact} to include mesh motion. We refer to \cref{fig:beltrami-solution} for a plot of the exact solution, and to \ref{app:beltrami} for a detailed description of the benchmark.  This test case is implemented in the \lifexcfd{} test \texttt{fluid\_dynamics\_cube}.
The default parameter file \texttt{lifex\_test\_fluid\_dynamics\_cube.prm} for this test can be generated by
\begin{lstlisting}
$ ./lifex_test_fluid_dynamics_cube -g
\end{lstlisting}
In all the tests in this section, we consider a hexahedral mesh and a second-order BDF scheme for temporal discretization.
Moreover, we consider the \textit{gradient-gradient} formulation of the diffusion term (see \eqref{eq:diffusion-term}).
In order to perform a convergence test in space, we exploit the capability of \lifex{} to refine the hexahedral meshes in a hierarchical way by splitting each hexahedron into eight hexahedra, halving the length of each edge.

These settings (as well as others specified below for each test) can be encoded by modifying the file \linebreak \texttt{lifex\_test\_fluid\_dynamics\_cube.prm} accordingly.
The simulations can be run by
\begin{lstlisting}
$ ./lifex_test_fluid_dynamics_cube -f lifex_test_fluid_dynamics_cube.prm
\end{lstlisting}

We set $T = \SI{1e-5}{\second}$, $\Delta t = \SI{1e-6}{\second}$, and perform a convergence test in space for $h = \num{0.085}$, \num{0.17}, \num{0.34}, \num{0.68}, \SI{1.36}{\meter} (corresponding to a number of mesh refinements equal to \num{7}, \num{6}, \num{5}, \num{4}, \num{3}, respectively), with $\mathbb{Q}^2-\mathbb{Q}^1$ finite elements and without any stabilization.
Moreover, we repeat the same test with $\mathbb{Q}^1-\mathbb{Q}^1$ finite elements and enabling SUPG-PSPG stabilization. In both cases, the domain displacement is computed using $\mathbb{Q}^1$ finite elements.

For both tests, we compute the following relative errors at the final simulation time between the numerical solution and the exact solution:
\begin{equation}
    e_{L^2}^{\vel} = \frac{\|\velh^{N_t} - \vel_\text{ex}\|_{L^2(\Omega)}}{\|\vel_\text{ex}\|_{L^2(\Omega)}} \qquad
    e_{H^1}^{\vel} = \frac{|\velh^{N_t} - \vel_\text{ex}|_{H^1(\Omega)}}{|\vel_\text{ex}|_{H^1(\Omega)}} \qquad
    e_{L^2}^{p} = \frac{\|\ph^{N_t} - p_\text{ex}\|_{L^2(\Omega)}}{\|p_\text{ex}\|_{L^2(\Omega)}}\;,
    \label{eq:errors}
\end{equation}
wherein the subscript ``ex'' denotes the exact solution (reported in full in \ref{app:beltrami}) and $\|\cdot\|_{L^2(\Omega)}$ and $|\cdot|_{H^1(\Omega)}$ denote the $L^2$ norm and $H^1$ seminorm, respectively.

\begin{figure}
    \centering
    \includegraphics{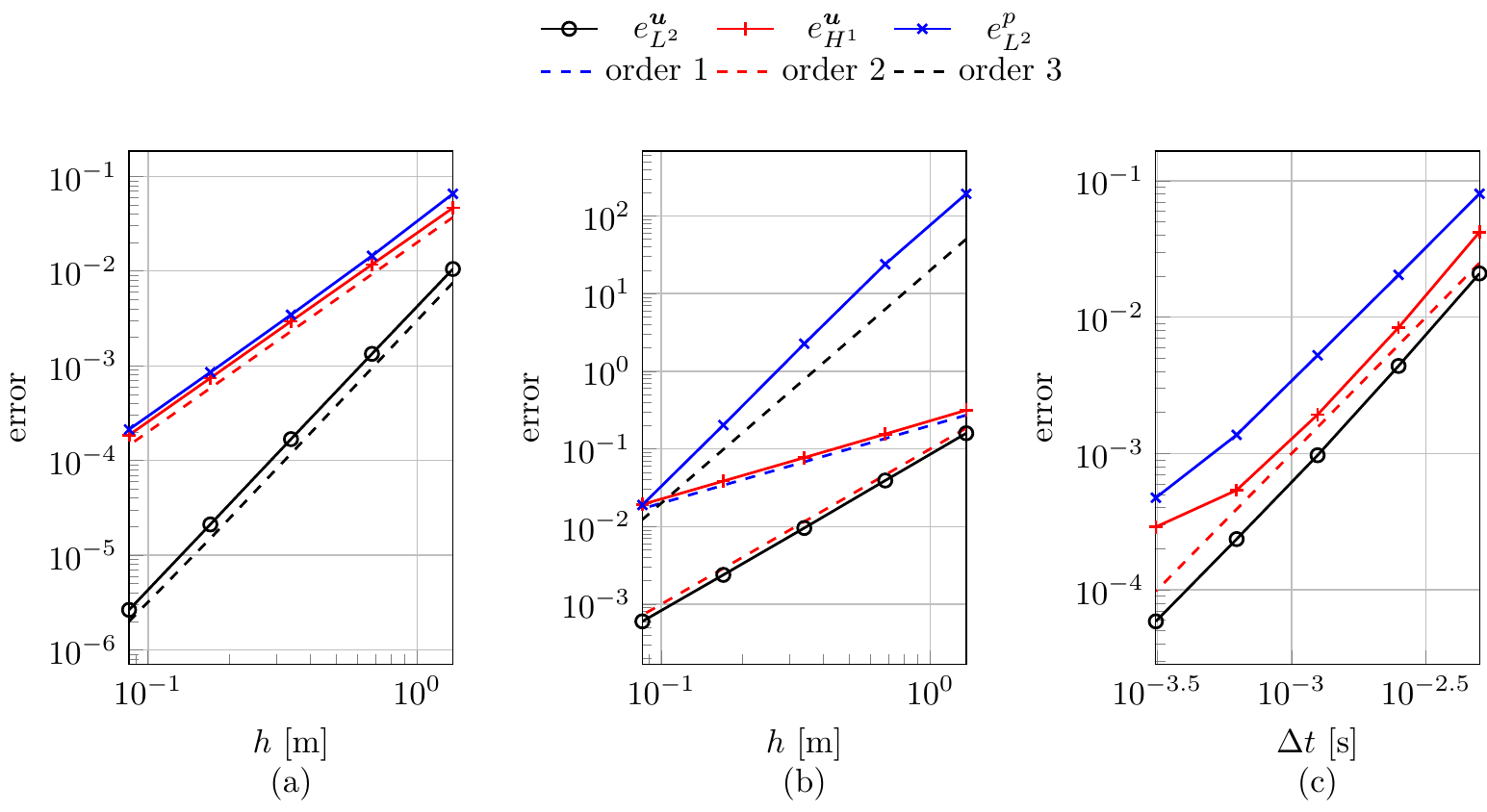}
    \caption{Test I. (a) Error against the mesh size $h$ with $\mathbb{Q}^2-\mathbb{Q}^1$ elements and without stabilization. (b) Error against the mesh size $h$ with $\mathbb{Q}^1-\mathbb{Q}^1$ elements with SUPG-PSPG stabilization. (c) Error against the time step size $\Delta t$ with $\mathbb{Q}^2-\mathbb{Q}^1$ elements and without stabilization. Dashed lines represent the theoretical convergence rates. Errors are measured at final time, according to \eqref{eq:errors}.}
    \label{fig:beltrami-convergence}
\end{figure}

The results are reported in Figures \ref{fig:beltrami-convergence}a and \ref{fig:beltrami-convergence}b, from which we can observe the expected convergence rates for the $L^2$ norm and $H^1$ seminorm of the velocity error and for the $L^2$ norm of the pressure error.
Indeed, from the theory of the finite element method, we expect the following orders of convergence on the relative errors when using a stable space discretization with $\mathbb Q^{m_{\vel}}-\mathbb Q^{m_p}$ finite elements and a second-order BDF time scheme \cite{girault2012finite}:
$e_{L^2}^{\vel}\sim C_1 h^{m_{\vel}+1} + C_2 \Delta t^2$,
$e_{H^1}^{\vel}\sim C_1 h^{m_{\vel}} + C_2 \Delta t^2$,
$e_{L^2}^{p}\sim C_1 h^{m_p+1} + C_2 \Delta t^2$.


We perform a convergence test in time, fixing the mesh size to $h = \SI{0.085}{\metre}$ (corresponding to \num{7} mesh refinements, \num{2097152} elements and \num{53070468} degrees of freedom) and progressively reducing the time step $\Delta t$ by a factor 0.5, from $\Delta t=\SI{5e-3}{\second}$ to $\Delta t=\SI{3.125e-4}{\second}$. We set $T = \SI{1e-2}{\second}$. The error, measured as in the previous test, tends to zero with the expected rate, as reported in \cref{fig:beltrami-convergence}c. A deflection in the trend of $e^{\vel}_{H^1}$ is observed for the finest time step, associated to the saturation of the error in space.

\begin{figure}
    \centering
    \includegraphics{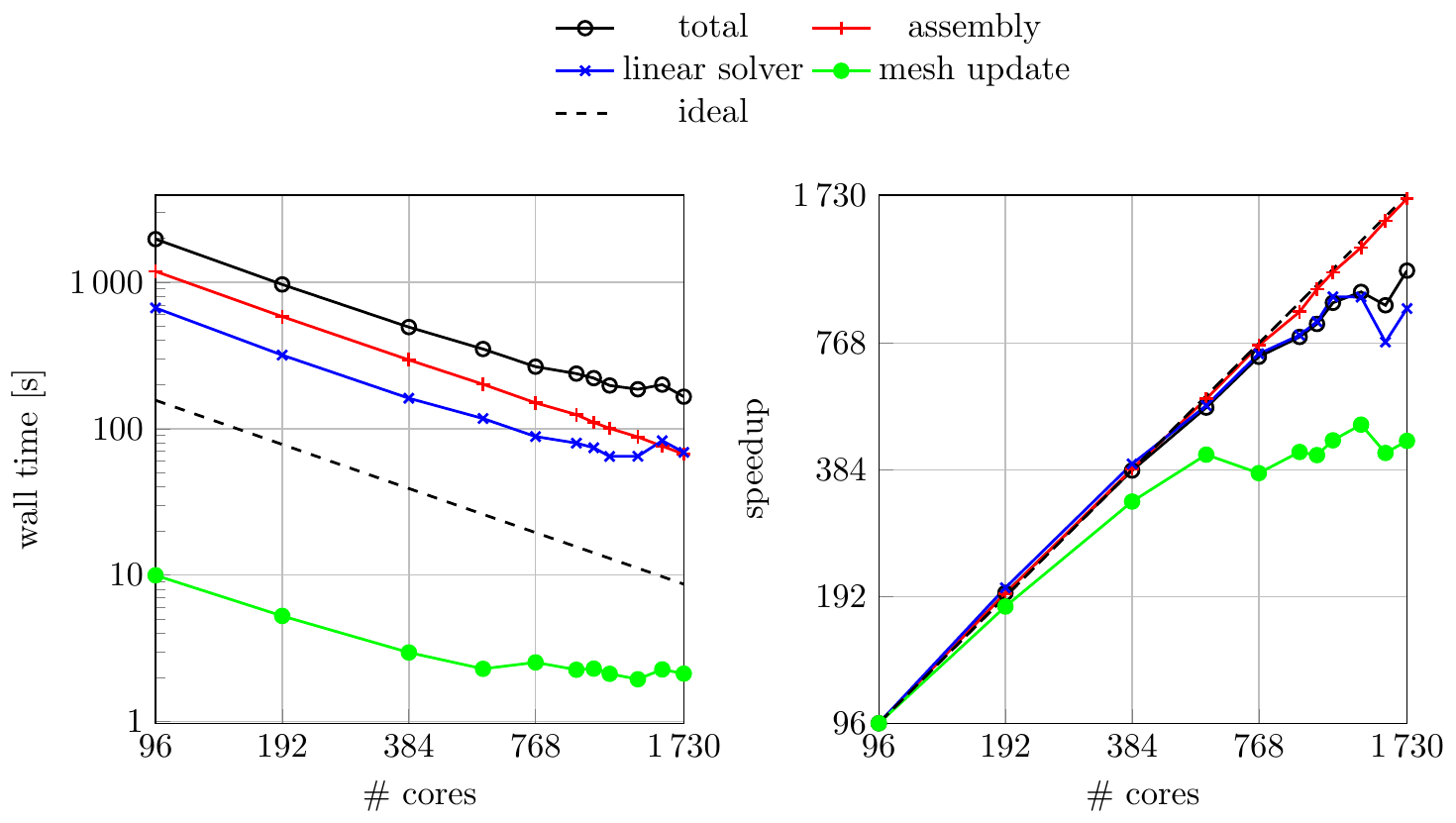}
    \caption{Test I. Strong scalability test. Wall times (left) and parallel speedup (right) varying the number of cores.}
    \label{fig:beltrami-scalability}
\end{figure}

Finally, we perform a strong scalability study for this test case, by measuring the computational times associated to the assembly of the linear system and its solution with a varying number of parallel cores. The simulation is run with mesh size $h = \SI{0.085}{\metre}$, corresponding to \num{2097152} elements and \num{53070468} degrees of freedom (and a number of mesh refinements equal to \num{7}). The test was run on the GALILEO100 supercomputer provided by the CINECA supercomputing center, Italy\footnote{528 computing nodes each 2 x CPU Intel CascadeLake 8260, with 24 cores each, 2.4 GHz, 384GB RAM. See \url{https://wiki.u-gov.it/confluence/display/SCAIUS/UG3.3\%3A+GALILEO100+UserGuide} for technical specifications.}. The results, plotted in \cref{fig:beltrami-scalability}, show that the solver scales linearly up to \num{768} cores (corresponding to approximately \num{70000} degrees of freedom per core), in agreement with \cite{deparis2014parallel, africa2022flexible}. In particular, the assembly stage scales almost perfectly, whereas the performance of the linear solver deteriorates starting from \num{768} cores. This is consistent with the observations of \cite{deparis2014parallel} on the performance of the aSIMPLE preconditioner. The mesh update step exhibits relatively poor scalability properties with respect to the other steps, due to the need to communicate the displacement of mesh nodes across processes. However, as shown in Figure 3a, this has a negligible impact on the overall scalability of the solver, since the mesh update step accounts for approximately \SI{1}{\percent} of the total wall time.

\subsection{Fluid dynamics in a compliant cylinder with a moving obstacle (Test II)}
\label{sec:cylinder}
We simulate the fluid dynamics in a compliant cylinder with a moving obstacle by using the \lifexcfd{} app introduced in \Cref{sec:quick-start}. A template parameter file for this setting (see also \cref{sec:quick-start}) can be obtained by running:
\begin{lstlisting}[language=bash]
$ ./lifex_fluid_dynamics -g -b "Inlet" "Outlet"
\end{lstlisting}
On the lateral wall of the cylinder, we prescribe no-slip boundary conditions. Therefore, we do not need to include it in the boundary label list (see below).
We consider a cylinder of length $L = \SI{0.1}{\meter}$ and radius $R = \SI{0.01}{\meter}$ and the hexahedral mesh provided by the \dealii{} mesh generator with 5 refinements (see \cref{tab:mesh} for the details on the mesh obtained). The following parameters are set in the parameter file\footnote{Each parameter is documented in the parameter file. In the paper, we omit the documentation for the sake of conciseness.}: 
\begin{lstlisting}[language=prm]
 subsection Mesh and space discretization
    set Mesh type             = Cylinder
    set Element type      	  = Hex
    set Number of refinements = 5
    
    subsection Cylinder
      set Radius = 0.01
      set Length = 0.1
    end
end
\end{lstlisting}
We set a no-slip boundary condition on the wall (tag 0), a Dirichlet condition on the inlet section (tag 1), and a homogeneous Neumann condition on the outlet section (tag 2). For the inlet section, we prescribe a parabolic profile, with a pulsatile temporal function with $g_\mathrm{min}= \SI{0}{\cubic\metre\per\second}$, $g_\mathrm{max}=\SI{2.5e-4}{\cubic\metre\per\second}$ and period $\mathcal T_\mathrm{inlet} = \SI{0.8}{\second}$ (see \Cref{sec:bcs-dirichlet}). 
With this choice of the input parameters, the resulting Reynolds number computed at the inlet section is:
\begin{equation}
    \mathrm{Re} = \frac{2 \, \rho \, g_\mathrm{max}}{\mu\, \pi\, R} = \num{4812}, 
\end{equation}
where we used, as characteristic velocity, the average velocity at the inlet section when the flowrate is maximum and the inlet section diameter as characteristic length. Furthermore, blood density and kinematic viscosity are set as \SI{1.06e3}{\kilo\gram\per\meter\cubed} and \SI{3.5e-3}{\pascal\second}, respectively.
\begin{lstlisting}[language=prm]
  subsection Boundary conditions
    set No-slip tags   = 0
    
    subsection Inlet
      set Tags                       = 1
      set Type of boundary condition = Dirichlet
      subsection Dirichlet
        set Time evolution     = Pulsatile
        set Space distribution = Parabolic
        subsection Pulsatile
          set Minimum value = 0.0
          set Maximum value = 2.5e-4
          set Period        = 0.8
        end
      end
    end

    subsection Outlet
      set Tags                       = 2
      set Type of boundary condition = Neumann
      subsection Neumann
        set Time evolution = Constant
        subsection Constant
          set Value = 0.0
        end
      end
    end
  end  
\end{lstlisting}
The domain boundaries are moving according to a displacement read from a \texttt{vtp} file.
This file contains the surface of the domain boundary and a sequence of 9 displacement fields, obtained as equispaced time samples (one every $\SI{0.1}{s}$) of the following displacement field:
\begin{equation*}
    \displ^{\partial \Omega} (\xhat, t) = C \, \frac{z(L-z)}{\sqrt{\widehat x^2+\widehat y^2}} \, \sin{\left (\frac{2\pi t}{T} \right )} \left( \widehat{x}, \, \widehat{y}, \, 0 \right)^T,  
\end{equation*}
with $C=0.015$, $\xhat = \left(\widehat x, \widehat y, \widehat z\right)^T$ and $T=\SI{0.8}{\second}$. 
These samples are interpolated in time by \lifexcfd{} with piecewise cubic splines.

Accordingly, the incompressible Navier-Stokes equations are solved in the ALE framework \eqref{eq:ns-ale-riis}, in combination with a lifting problem \eqref{eq:lifting} for the boundary displacement $\displ^{\partial\Omega}$.
We use the harmonic lifting, solved with the CG method with AMG preconditioning (see \cref{sec:lifting}).
\begin{lstlisting}[language=prm]
  subsection Arbitrary Lagrangian Eulerian
    set Active                        = true
    set Import displacement from file = true
    
    subsection Input file
      set Boundary displacements filename      = displacement_cylinder.vtp
      set Boundary displacement field basename = d
      set Time subintervals                    = 8
      set Time subinterval duration            = 0.1
      set Interpolation mode                   = Splines
    end
    
    subsection Lifting
      set Lifting operator = Harmonic
      set Tags Dirichlet   = 0, 1, 2
    end
  end  
\end{lstlisting}

In the domain, a moving surface obstacle representing an idealized valve leaflet is immersed by the RIIS method.
The closed configuration of the valve and its displacement field \texttt{openingField} are stored in a file \texttt{cylinder\_plane\_closed.vtp}.
The parameters chosen for the RIIS method are $\varepsilon=\SI{3}{\milli\meter}, R=\SI{1e4}{\kilo\gram\per\meter\per\second}$. The valve starts in closed configuration, opens for $t \in [0.15, 0.25]\si{\second}$, stays open for $t \in [0.25, 0.55]\si{\second}$, then closes during the interval $t \in [0.55, 0.65]\si{\second}$ and stays closed from there on.
The valve follows the displacement of the domain, whereas we use a quasi-static approach for its opening/closing, that is we set $\bf u_\Sigma\equiv\bf 0$ in the RIIS term \eqref{eq:riis} \cite{fedele2017patient, fumagalli2020image}.
\begin{lstlisting}[language=prm,escapechar={|}]
  subsection Resistive Immersed Implicit Surface
    set Active                      = true
    set Use surface velocity        = false
    set Surface labels              = surface
    set Immersed surfaces basenames = cylinder_plane_closed
    set Move surfaces with ALE      = true
    set Displacement names          = openingField
    set Epsilons                    = 0.003
    set Resistances                 = 10000

    subsection Displacement law
      set Displacement laws = Prescribed

      subsection Prescribed
        set First ramp: start time  = 0.15
        set First ramp: |end| time    = 0.25
        set Second ramp: start time = 0.55
        set Second ramp: |end| time   = 0.65
      end
    end
  end
\end{lstlisting}

For the solution of the problem, we use a semi-implicit treatment of the advection term, thus the problem is linear.
To prescribe a single Newton iteration in the solution, the following parameters are used:
\begin{lstlisting}[language=prm]
  subsection Time solver
    set BDF order               = 1
    set Non-linearity treatment = Semi-implicit
    set Initial time            = 0
    set Final time              = 0.8
    set Time step               = 1e-4
  end
  
  subsection Non-linear solver
    set Linearized = true
  end
\end{lstlisting}
Notice that the section \texttt{Time solver} also allows to setup the simulation to run from $t=\SI{0}{\second}$ to $T=\SI{0.8}{\second}$, with $\Delta t=\SI{1e-4}{\second}$, using a semi-implicit Euler time scheme (BDF1).

\begin{figure}
    \centering
    \includegraphics[width=\textwidth]{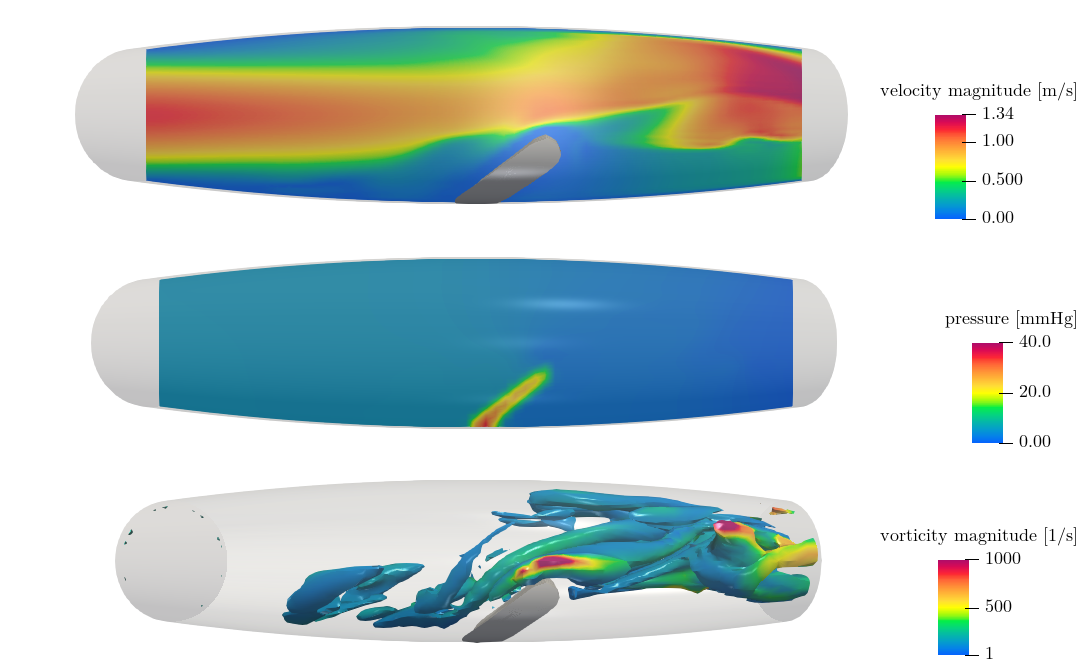}
    \caption{Test II. Flow results at $t=\SI{0.26}{\second}$: velocity magnitude (top) and pressure (center) distributions on a median slice, and Q-criterion contours (bottom, $Q=\SI{1000}{\per\square\second}$) colored by the vorticity magnitude. In grey, the region occupied by the valve.}
    \label{fig:cylinder}
\end{figure}

This test case, featuring an ALE approach with the introduction of an immersed valve, serves as a prototype for a cardiovascular CFD simulation on an idealized vessel. The results of the simulation at $t=\SI{0.26}{\second}$ are displayed in \cref{fig:cylinder}.
At this time, the valve is in its open position: in the figure, we report in grey the support of the Dirac delta of the RIIS term:
\[
\{\bm{x}\in\Omega_t \colon |\varphi(\bm{x})|<\varepsilon\}.
\]
As the flow is introduced from the left into the computational domain, the presence of the immersed valve exerts a noticeable influence, causing a deflection in the oncoming flow. This dynamic interaction leads to the generation of a majority of the coherent vorticity structures, as strikingly revealed by the Q-criterion contours. These vorticity structures are not only visually captivating but also testify the intricate interplay between the fluid flow and the valve. The pressure plot intentionally excludes the representation of the valve. This omission allows for a clear and unobstructed perspective on the development of a high pressure jump, which is prominent in the immediate vicinity of the valve. This region effectively transforms the valve into an immersed obstacle with a significant impact on the flow dynamics.

\subsection{Application to a vascular case (Test III)}
\label{sec:vessel}

\begin{figure}
    \centering
    \includegraphics[width=0.65\textwidth]{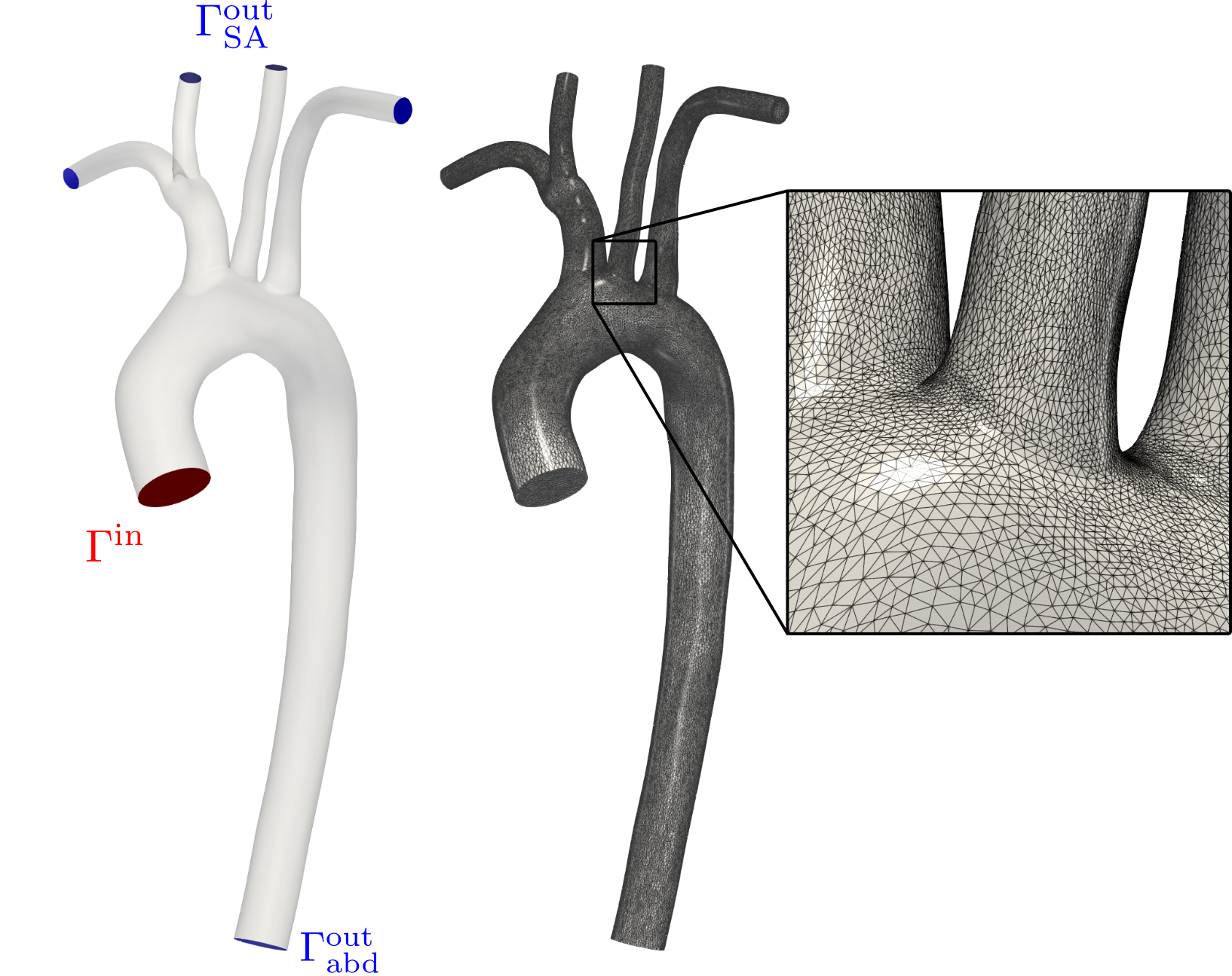}
    \caption{Test III. Computational domain and mesh, obtained from the Vascular Model Repository \cite{wilson2013vascular}}
    \label{fig:aorta-domain}
\end{figure}

We simulate the hemodynamics in a portion of aorta from a healthy 11-year-old male, also simulated in \cite{ladisa2011computational}. The mesh, shown in \cref{fig:aorta-domain}, and the associated inlet boundary data were obtained from the Vascular Model Repository \cite{wilson2013vascular}. The mesh was preprocessed with \texttt{vmtk} \cite{vmtk} to convert it to the \texttt{msh} format supported by \lifexcfd{}.

The domain features one inlet, corresponding to the aortic root section, and four outlets, corresponding to supra-aortic arteries and abdominal aorta. The domain does not move (i.e. we neglect the compliance of the walls) and no valves are present (i.e. ALE and RIIS are disabled). We impose Dirichlet boundary conditions on the inlet section $\Gamma^\text{in}$, with a parabolic velocity profile and a flow rate obtained by time interpolation of a given datum (see \cref{sec:bcs-dirichlet}). The flow rate over time comes from experimental measurements, and is characterized by a strong ejection phase during ventricular systole, followed by a diastolic phase with nearly zero inlet flow. On the outlet sections $\Gamma^\text{out}$, we impose resistance boundary conditions (\cref{sec:bcs-resistance}), setting $C_\mathrm{R} = \SI{1e9}{\kilo\gram\per\second\per\metre\tothe{4}}$ on the supra-aortic outlets $\Gamma^\text{out}_\text{SA}$ and $C_\mathrm{R} = \SI{3.5e7}{\kilo\gram\per\second\per\metre\tothe{4}}$ on the abdominal outlet $\Gamma^\text{out}_\text{abd}$. Finally, on the external walls we impose no-slip boundary conditions.

With this choice of the input parameters, the resulting Reynolds number computed at the inlet section is $\mathrm{Re} = \num{5260}$, where we used as characteristic velocity the space-averaged velocity at the inlet section when the flowrate is maximum, and as characteristic length the inlet section diameter.

A template parameter file for this setting (see also \cref{sec:quick-start}) can be obtained by running:
\begin{lstlisting}[language=bash]
$ ./lifex_fluid_dynamics -g \ 
        -b "Aortic root" "Right subclavian" \
           "Right common carotid" "Left common carotid" \ 
           "Left subclavian" "Abdominal aorta"
\end{lstlisting}

\begin{figure}
    \centering
    \begin{subfigure}[b]{0.5\textwidth}
        \centering
        \includegraphics{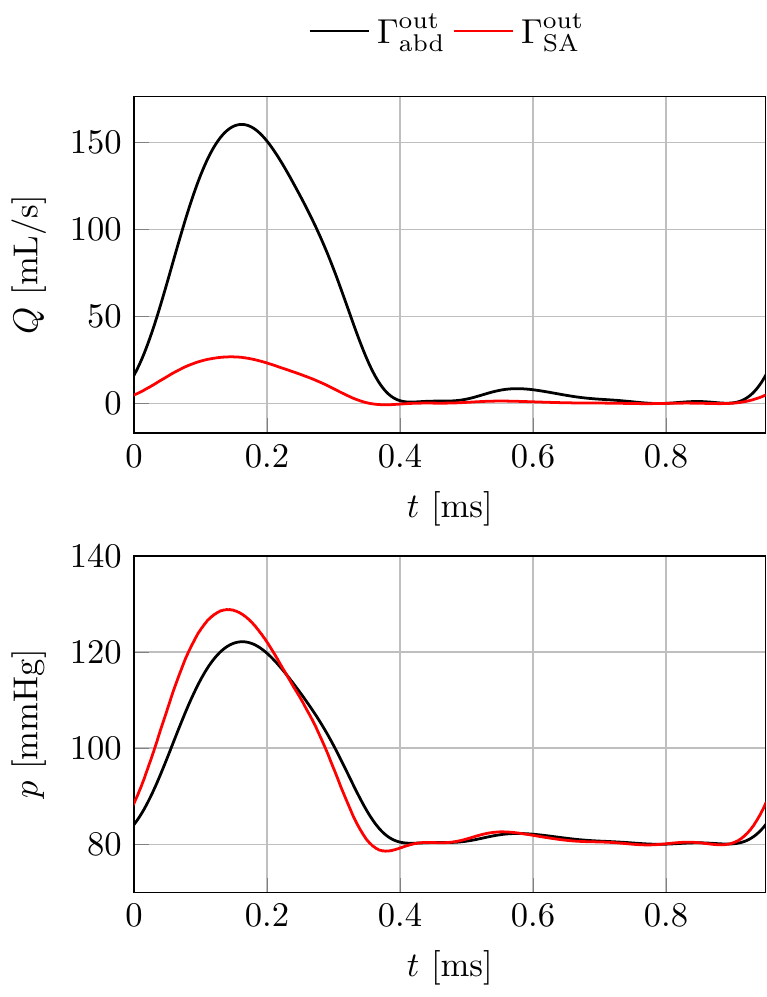}
        \caption{}
        \label{fig:aorta-outlet}
    \end{subfigure}
    \hspace{0.05\textwidth}
    \begin{subfigure}[b]{0.4\textwidth}
        \centering
        \includegraphics[width=\textwidth]{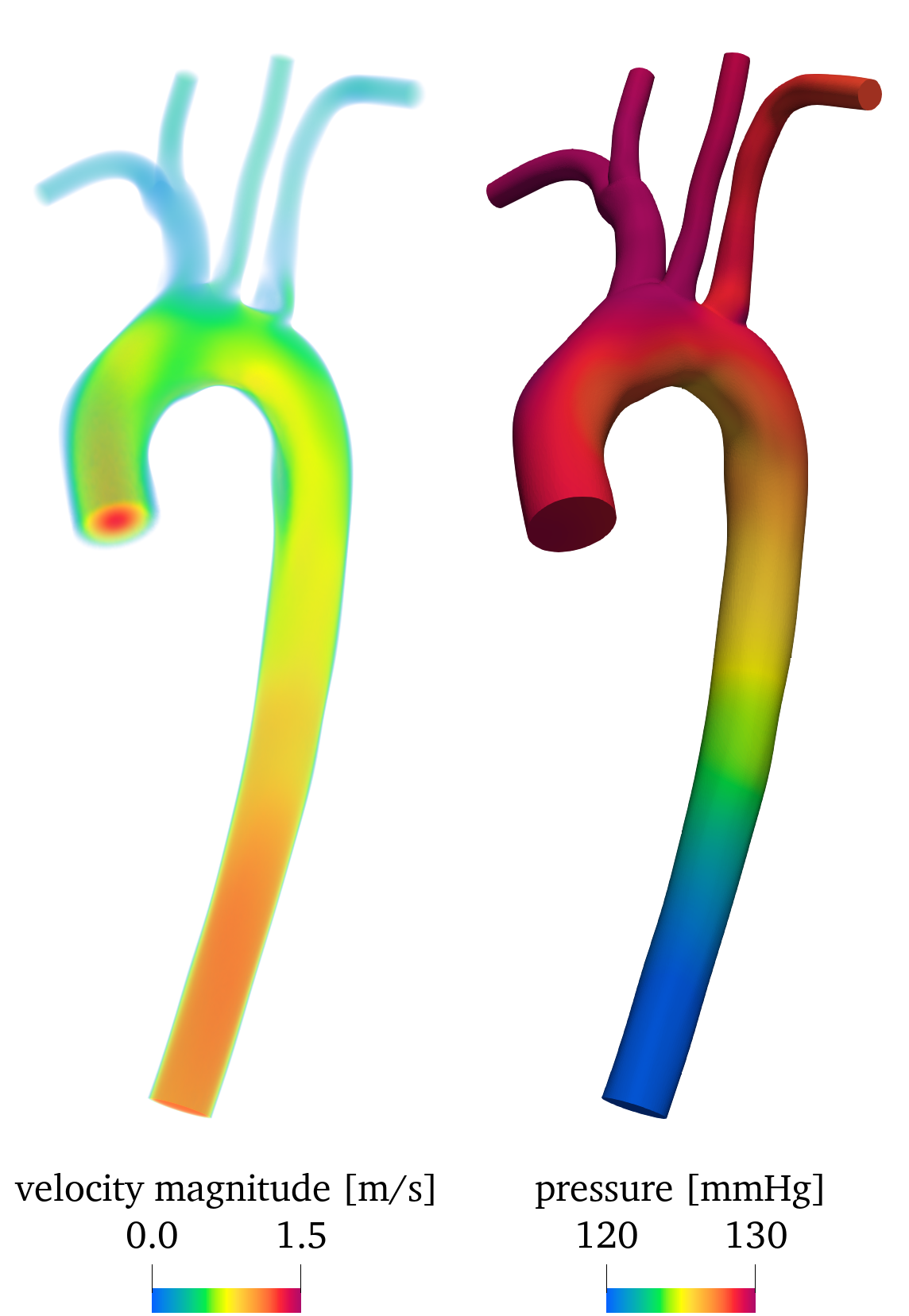}
        \caption{}
    \end{subfigure}
    
    \caption{Test III. (a) Flowrate and average pressure on the supra-aortic and abdominal outlets during the last heartbeat of the simulation. (b) Volume rendering of the blood velocity magnitude and pressure at time $t = \SI{0.15}{\second}$ of the last simulated cycle.}
    \label{fig:aorta-results}
\end{figure}

We simulate three heartbeats of period $t_\text{hb} = \SI{0.95}{\second}$ ($T = 3t_\text{hb} = \SI{3.8}{\second}$), with a time step $\Delta t = \SI{0.001}{\second}$. We report in \cref{tab:mesh} details about the mesh used for this test case. We used $\mathbb{P}^1-\mathbb{P}^1$ finite elements, with SUPG-PSPG stabilization, and the BDF1 time discretization method with the semi-implicit formulation for the advection term. The simulation was run on \num{92} parallel cores, endowed with Intel Xeon Platinum 8160 CPUs. The total wall time in this setting was \SI{6.3}{\hour}. \Cref{fig:aorta-results} displays numerical results for this test case, which are in qualitative accordance with \cite{ladisa2011computational}. The solver can also be configured to compute post-processing quantities such as flow rate or average pressure on specific portions of the boundary. As an example, we report the average pressure and flow rate through $\Gamma^\text{out}_\text{SA}$ and $\Gamma^\text{out}_\text{abd}$ in \cref{fig:aorta-outlet}.

\begin{table}
    \centering
    \small
    \begin{tabular}{l c c c c c c c c c}
        \toprule
        \textbf{Test case} & \textbf{Type} & \textbf{\# elements} & \textbf{\# vertices} & $h_{\min}$  & $h_{\mathrm{avg}}$ & $h_{\max}$ & $q_{\min}$ & $q_\mathrm{avg}$ & $q_\mathrm{max}$  \\
        &  &  &  & [\si{\milli\meter}]  & [\si{\milli\meter}] & [\si{\milli\meter}]  \\
        \midrule
        I (Beltrami flow, \cref{sec:software-verification}) & Hex & - & - & - & - & - & \num{1} & \num{1} & \num{1} \\
        II (cylinder, \cref{sec:cylinder}) & Hex & 81920 & 85345 & 1.64 & 1.75 & 1.91 & \num{3.11} & \num{4.02} & \num{5.38} \\
        III (aorta, \cref{sec:vessel}) & Tet & \num{2915690} & \num{543089} & \num{0.08} & \num{0.8} & \num{2.9} & \num{1.04} & \num{2.96} & \num{16.83} \\
        IV (left atrium, \cref{sec:left-atrium}) & Tet & \num{2438278} & \num{389484} & \num{0.28} & \num{0.8} & \num{1.3} & \num{1.00} & \num{1.54} & \num{3.08} \\
        \multirow{2}{*}{V (Taylor-Green, \cref{sec:tgv})} & Hex & \num{32}\textsuperscript{3} & \num{35937} & \num{0.34} & \num{0.34} & \num{0.34}& \num{1} & \num{1} & \num{1} \\
        & Hex & \num{64}\textsuperscript{3} & \num{274625} & \num{0.17} & \num{0.17} & \num{0.17} & \num{1} & \num{1} & \num{1} \\
        \bottomrule
    \end{tabular}
    \caption{Mesh type (Hex: hexaedral, Tet: tetrahedral), number of mesh elements, number of mesh vertices, minimum, average and maximum mesh size and minimum, average and maximum element edge ratio $q$ for the test cases presented. The edge ratio $q$ is the ratio between the longest and shortest edge of a mesh element. Details on Test I are given in \cref{sec:software-verification} since a mesh refinement study is carried out.}
    \label{tab:mesh}
\end{table}

\subsection{Application to a cardiac case (Test IV)}
\label{sec:left-atrium}
\begin{figure}[t]
    \centering
    \includegraphics[trim={0 0 0 0},clip,width = \textwidth]{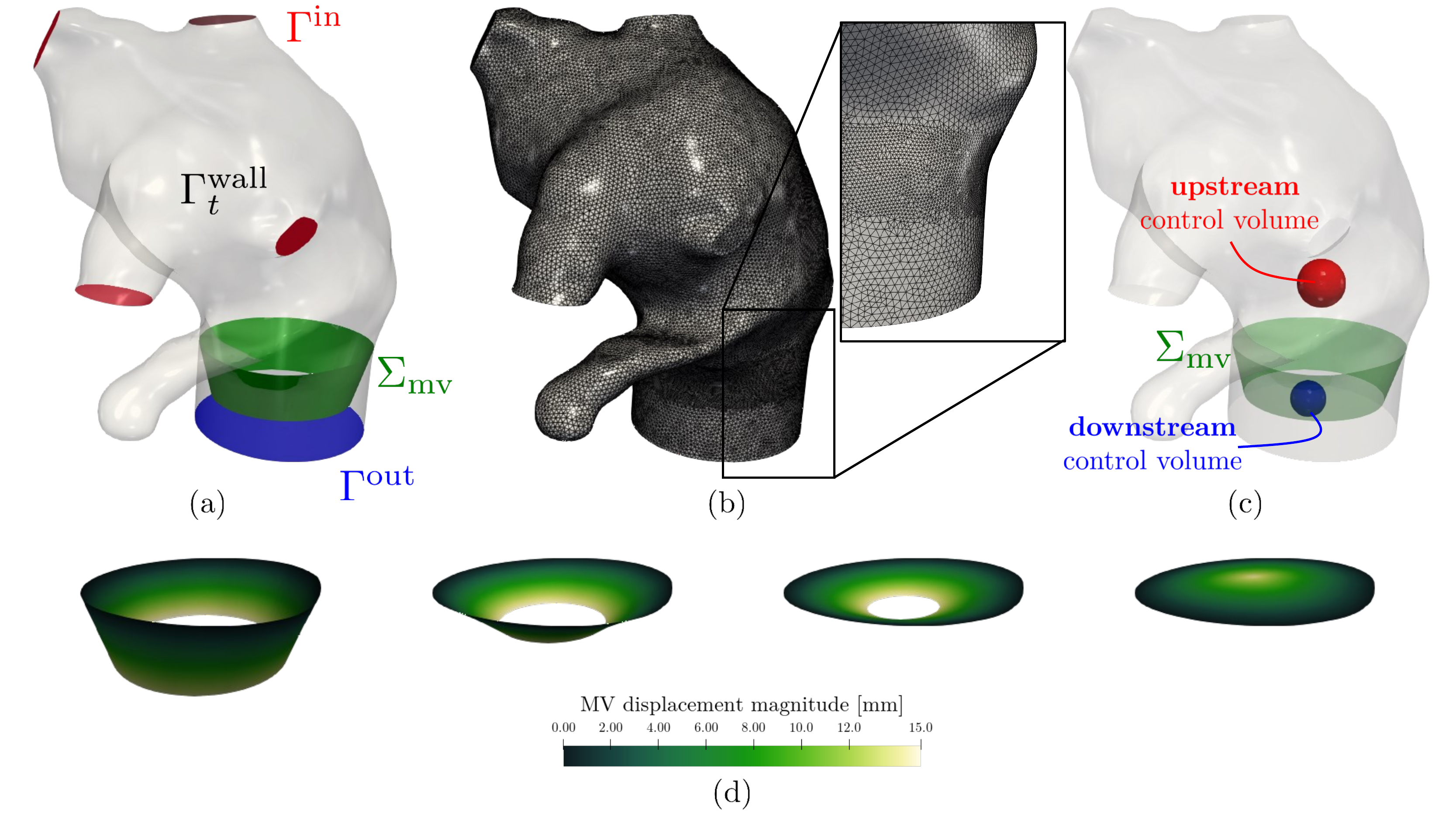}
    \caption{Test IV. (a): left atrium geometry, with boundary portions; (b) tethraedral mesh refined on the proximity of the mitral valve; (c) control volumes upstream the mitral valve; (d) mitral valve warped by its displacement (from open to closed configuration).}
    \label{fig:la-geometry-mesh}
\end{figure}

We simulate the hemodynamics in a patient-specific left atrium geometry. We consider the atrial geometry marked with ID 7 available in the public cohort \cite{niederer:geom, roney:geom}. As we display in \Cref{fig:la-geometry-mesh}a, the left atrium boundary is made of four pulmonary veins inlet sections ($\Gamma^\mathrm{in}$), an outlet section downstream of the mitral valve ($\Gamma^\mathrm{out}$), and the endocardial wall ($\Gamma^\mathrm{wall}_t$). 

We generate the tetrahedral mesh of the left atrium in \Cref{fig:la-geometry-mesh}b using \texttt{vmtk}~\cite{vmtk} with the methods and tools discussed in~\cite{fedele2021polygonal}. The mesh is locally refined to capture the mitral valve leaflets as described by the RIIS method.
Indeed, the application of the RIIS method requires the mesh to be sufficiently fine to accurately represent the smoothed Dirac delta function $\delta_{\Sigma,\varepsilon}$ of thickness $2\varepsilon$. As a rule of thumb, this requirement is fulfilled if the local mesh size $h$ satisfies $\varepsilon\geq 1.5h$, as shown in \cite{fedele2017patient}.
Details on the mesh we used are provided in \Cref{tab:mesh}. 

Since the patient-specific geometry of the mitral valve is not available, we generate an idealized valve geometry $\Sigma_\mathrm{mv}$, and the corresponding displacement, displayed in Figures~\ref{fig:la-geometry-mesh}a, \ref{fig:la-geometry-mesh}c and \ref{fig:la-geometry-mesh}d. We open and close the valve according to pressure jump conditions. Specifically, we monitor the pressures in control volumes located upstream and downstream the valve, as shown in \cref{fig:la-geometry-mesh}c. The mitral valve opens when the atrial pressure overcomes the ventricular one. Viceversa, the valve closes when the pressure drop inverts its sign. Furthermore, we do not open and close the valve instantaneously, but we consider times coming from medical literature: the valve opens in \SI{20}{\milli\second} \cite{MV:open} and closes in \SI{60}{\milli\second} \cite{MV:close}. The opening and closing ramps follow the \textit{cosinusoidal-exponential} law used in \cite{bucelli2022mathematical}. In Figures \ref{fig:la-plots}b and \ref{fig:la-plots}d, we report the pressure computed in the upstream and downstream control volumes and the opening coefficient of the mitral valve, respectively. Furthermore, we set a resistance coefficient $R_\mathrm{MV} = \SI{1e4}{\kilo\gram\per\metre\per\second}$, such that the valve is sufficiently impervious, and we set $\varepsilon_\mathrm{MV} = \SI{0.7}{\milli\metre}$ by averaging the values of the mitral leaflet thicknesses reported in~\cite{CRAWFORD20011419}.

As boundary data, we prescribe Neumann boundary conditions on the inlet pulmonary veins sections and on the section downwind the mitral valve. We use the boundary data that we obtained in \cite{corti2022impact} by calibrating a 0D circulation model of the whole cardiovascular system \cite{regazzoni2022cardiac}. On $\Gamma^{\mathrm{in}}$, we prescribe the pulmonary venous pressure, whereas we set the left ventricular pressure on  $\Gamma^{\mathrm{out}}$. Inlet and outlet prescribed pressures are displayed in \Cref{fig:la-plots}c. On the endocardial wall, we set a no-slip condition by prescribing the displacement field that we computed in \cite{corti2022impact}. We refer the reader to \cite{corti2022impact} for additional details on the procedure we use to obtain boundary pressures and displacement.

We simulate 3 heartbeats of period $t_\mathrm{hb}=1.0$ s ($T = 3t_\mathrm{hb}=3.0$ s), with a time step size $\dt = \num{5e-4}\; \si{\second}$. We use $\mathbb P^1$-$\mathbb P^1$ FE spaces, BDF1 as time integration scheme, semi-implicit treatment of nonlinearities, and the VMS-LES methods. Numerical simulations are run in parallel on the GALILEO100 supercomputer, using 240 cores. The wall time for this simulation was about \SI{22}{\hour}.

The simulation is carried out via the \lifexcfd{} app. Since we are setting the same boundary condition on all the inlet sections, the parameter file for this setting (see also \cref{sec:quick-start}) can be obtained by running:
\begin{lstlisting}[language=bash]
$ ./lifex_fluid_dynamics -g -b "Pulmonary veins" "Mitral valve"
\end{lstlisting}

\begin{figure}[t]
    \centering
    \includegraphics{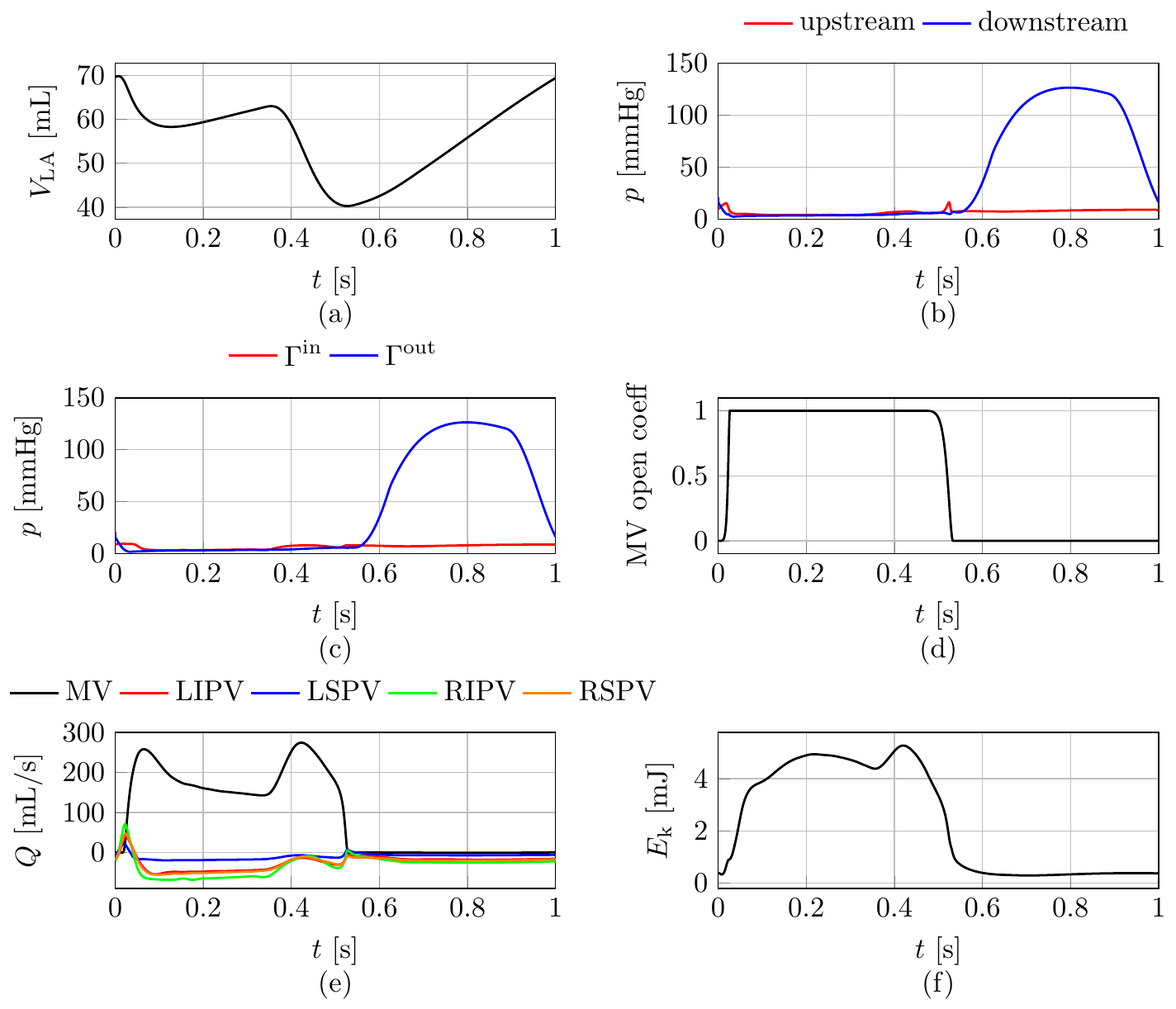}
    \caption{Test IV. Plots of left atrial CFD simulation results. (a) Left atrial volume versus time; (b) pressure in upstream and downstream control volumes (the control volumes are reported in \Cref{fig:la-geometry-mesh}c); (c) prescribed pressure at inlet and outlet sections; (d) mitral valve opening coefficient; (e) flowrates at inlets and outlet sections versus time; (f) kinetic energy versus time.}
    \label{fig:la-plots}
\end{figure}

The Reynolds number computed at the outlet section with the spatial-averaged velocity when the flowrate is maximum and the outlet section diameter as characteristic length is equal to \num{3792}.
In \cref{fig:la-plots}, we report some representative output of a \lifexcfd{} CFD cardiac simulation. We highlight that all the reported quantities are directly computed in \lifexcfd{} and exported to a CSV file.

We report solutions on the last simulated heartbeat and shifted in the time domain $(0, t_\mathrm{hb})$. For instance, we show the left atrial volume versus time in \cref{fig:la-plots}a. In \cref{fig:la-plots}e, we display the flow rates through each pulmonary vein and at the outlet section (i.e. downstream of the mitral valve), and \cref{fig:la-plots}f shows the kinetic energy against time. 

\begin{figure}[t]
     \centering
     \begin{subfigure}[b]{0.49\textwidth}
         \centering
         \includegraphics[width=\textwidth]{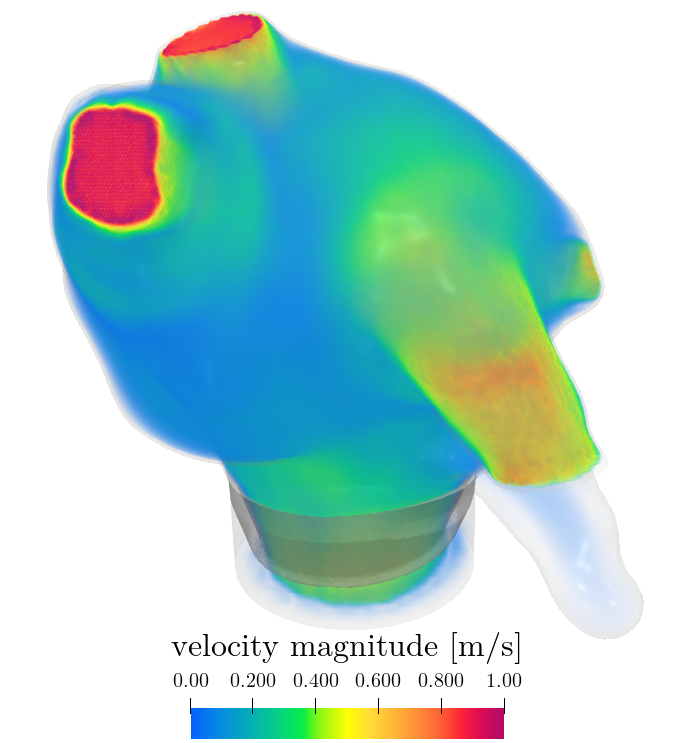}
         \caption{velocity}
         \label{fig:la-results-3D-velocity}
     \end{subfigure}
     \hfill
     \begin{subfigure}[b]{0.49\textwidth}
         \centering
         \includegraphics[width=\textwidth]{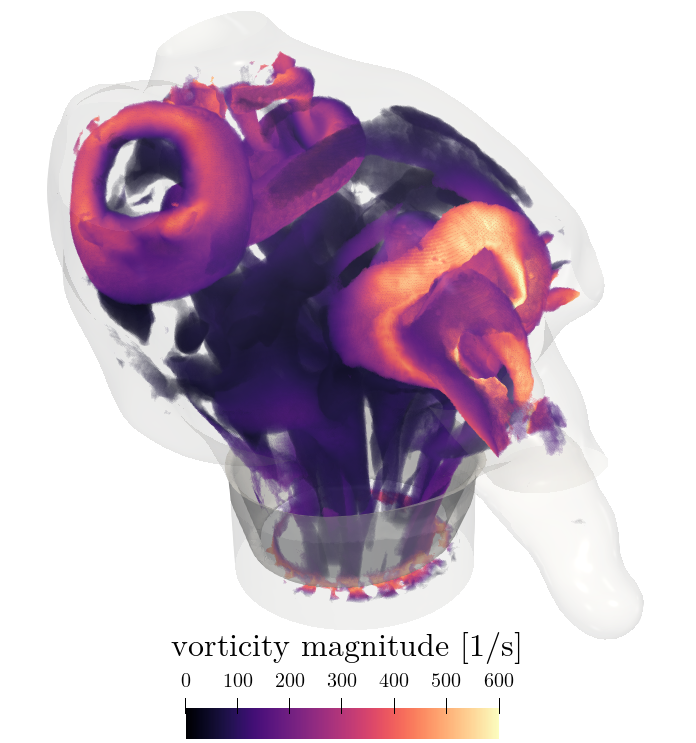}
         \caption{Q criterion}
         \label{fig:la-results-3D-qcriterion}
     \end{subfigure}
        \caption{Test IV. 3D visualizations of left atrium simulation at the early-wave peak. (a) volume rendering of velocity magnitude; (b) volume rendering of Q criterion ($Q = \SI{50}{\per\square\second}$) coloured according to vorticity magnitude.}
        \label{fig:la-results-3D}
\end{figure}

Finally, in \cref{fig:la-results-3D} we display some 3D representations of the CFD atrial simulation: we report the volume rendering of the velocity magnitude and the Q-criterion at the A-wave peak, when the incoming jets impact in the middle of the chamber, along with four ring vortices coming from the pulmonary veins.

We refer the interested reader to \cite{zingaro2022geometric,zingaro2023electromechanics,fumagalli2022image,marcinno2022computational} for a more detailed assessment of the results provided by \lifexcfd{} in realistic and patient-specific settings.

\subsection{A turbulent benchmark: the Taylor Green Vortex benchmark (Test V)}
\label{sec:tgv}

We show the capability of the code to capture turbulent flows by simulating the 3D Taylor-Green Vortex benchmark problem \cite{brachet1983small}. We set the problem in a cubic domain $\Omega = [0, 2\pi L]^3$ with final time $T=\SI{10}{\second}$. We consider a non-null initial condition for the velocity
\begin{equation*}
    \bm u_0 = U_0 
    \begin{bmatrix}
        \sin \left (\frac{x}{L}\right) \cos \left (\frac{y}{L}\right) \cos \left (\frac{z}{L}\right)  \\
        -\cos \left (\frac{x}{L}\right) \sin \left (\frac{y}{L}\right) \cos \left (\frac{z}{L}\right)  \\
        0
    \end{bmatrix} \; \text{in } \Omega \times \{0\},
\end{equation*}
and we set periodic boundary conditions on all the faces of the cube (see \Cref{sec:bcs-periodic}). $L = \SI{1}{m}$ and $U_0 = \SI{1}{\meter\per\second}$ are the characteristic length and characteristic velocity, respectively. Following \cite{brachet1983small}, we consider a Reynolds number equal to $\mathrm{Re} = \num{1600}$ by setting unity density and $\mu = 1/\mathrm{Re}$. We consider two mesh levels made of $32^3$ and $64^3$ elements on structured grids, reproducible by setting the parameter \texttt|Mesh and space discretization / Number of refinements| equal to 5 and 6, respectively.
Further details on these meshes are reported in \Cref{tab:mesh}.
We use quadratic FEs for velocity and pressure ($\mathbb Q^2-\mathbb Q^2$), time-step size equal to $\SI{2e-3}{\second}$ and the VMS-LES method \cite{bazilevs2007variational, forti2015semi} both to stabilize the problem and to model turbulence (see \Cref{sec:methods}).

\begin{figure}[t]
    \centering
    \includegraphics{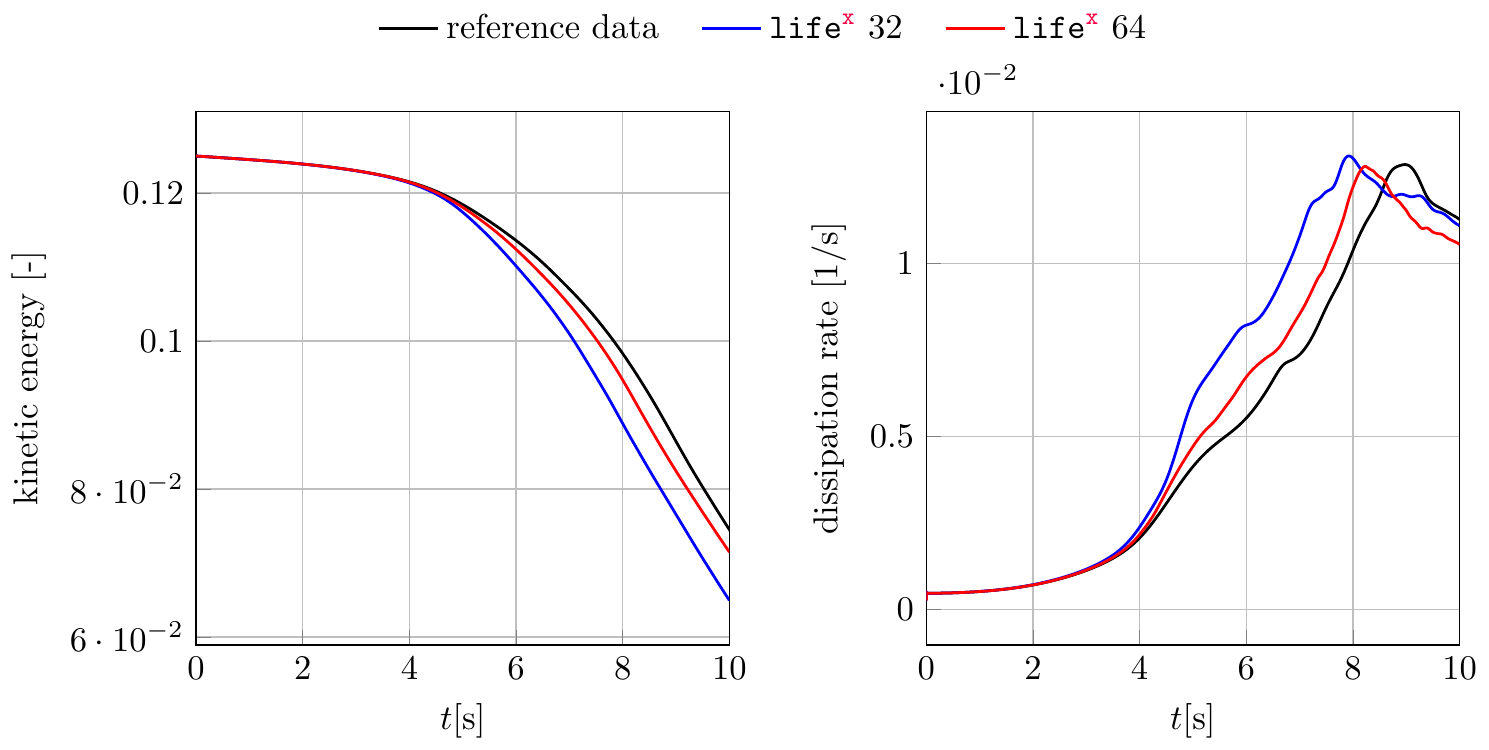}
    \caption{Test V: Numerical results of the Taylor-Green Vortex problem. Comparison between results obtained with \lifexcfd{} on two different grid levels and the DNS (reference) data by \cite{brachet1983small, tgv-dns-data}. Left: kinetic energy; right: dissipation rate.}
    \label{fig:tgv}
\end{figure}

\Cref{fig:tgv} shows a comparison between normalized kinetic energy $e(t)$ and dissipation rate $\epsilon(t)$ for the two grid levels against DNS reference data provided in \cite{brachet1983small, tgv-dns-data}. These quantities are computed as
\begin{equation*}
    e(t) = \frac{1}{\rho |\Omega|} \int_{\Omega} \frac{1}{2} \rho \bm u \cdot \bm u, \qquad \epsilon(t) = - \dv{e(t)}{t}.
\end{equation*}
Our results show that, as the grid is refined, the solution obtained in \lifexcfd{} gets closer to the reference solution, both in terms of kinetic energy and dissipation rate, showing how the solver is able to accurately capture turbulent flows. 

\section{Conclusions}
\label{sec:conclusions}

We presented the open-source \lifexcfd{} solver aimed at cardiovascular CFD simulations. The solver employs an ALE formulation to solve the incompressible Navier-Stokes equations for both laminar and turbulent regimes. It incorporates the SUPG-PSPG and VMS-LES stabilization methods, with the latter also serving as a turbulence model. This enables the solver to accurately model the transition-to-turbulence regime present in cardiovascular blood flow, as demonstrated by our results of the Taylor-Green Vortex benchmark problem. Cardiac valves can be included by means of the \textit{Resistive Immersed Implicit Surface} method. Several types of flow and stress boundary conditions are also supported, and exposed to the user in a friendly and highly customizable way. 

The numerical schemes incorporated into the code allow to discretize the equations in time using \textit{Backward Differentiation Formulas} (BDF) of order $\sigma_\bdf = 1, 2, 3$ and in space using Lagrangian Finite Elements (FE) of order \num{1} and \num{2} on tetrahedral meshes, and of arbitrary degree on hexahedral meshes. \textit{Fully implicit} and \textit{semi-implicit} formulations of the fluid dynamics problem are available. Thanks to automatic differentiation, the implementation of this complex and versatile formulation has become both effortless and efficient, making it an invaluable tool for new developers to further research. Furthermore, the user can also fine-tune parameters for the linear solvers and preconditioners.

We describe the \lifexcfd{} implementation details and provide a step-by-step guide for running simulations, showcasing the solver's user-friendliness and versatility in simulating different cardiovascular applications. To demonstrate the capabilities of the solver, we present multiple representative examples, such as simulating vascular flow in an idealized cylinder, the hemodynamics in a patient-specific portion of the aorta, and the left atrial hemodynamics in a patient-specific geometry. The solver's accuracy is verified through convergence analyses on the Beltrami flow benchmark problem. We also demonstrate its parallel performance, achieving a nearly-ideal speedup up to \num{768} processors. Overall, the tests and the results presented confirmed the effectiveness and efficiency of \lifexcfd{} for simulating complex cardiovascular flow problems.

Our CFD solver is designed to be user-friendly, efficient, and accurate, while also providing a flexible and extensible platform for future development and research in cardiovascular modeling and simulation. The open-source nature of \lifexcfd{} allows for collaboration and contributions from the broader scientific community, and we hope that it will serve as a valuable companion for researchers and practitioners working in the field of cardiovascular fluid dynamics, ultimately leading to improved diagnosis and treatment of cardiovascular diseases.

The impact and wide applicability of \lifexcfd{} are demonstrated by the high number of journal articles \cite{bennati2023turbulence, corti2022impact, zingaro2022geometric, bucelli2022partitioned, bucelli2022mathematical, bucelli2022stable, di2022prediction, fumagalli2022image, marcinno2022computational, bennati2023image, zingaro2022modeling, zingaro2023comprehensive} and preprints \cite{zingaro2023electromechanics, fumagalli2021reduced,renzi2023accurate} that have already used it in a diverse set of applications.
In particular, \lifexcfd{} allows to perform segregated (or one-way coupled) fluid-structure interaction simulations, where the boundary displacement is imported from the output of independent (electro)mechanical simulations \cite{zingaro2022geometric,zingaro2023electromechanics,zingaro2023comprehensive}.

Considering that CFD cardiovascular simulations often require the coupling with lumped-parameter models describing the outer part of the domain of interest, we plan to release the coupling with both Windkessel models and 0D closed-loop circulation models, as we do for instance in \cite{zingaro2022geometric, bucelli2022mathematical, marcinno2022computational, zingaro2023electromechanics}.
In addition, future developments pursued regarding \lifexcfd{} include 
the extension to more sophisticated models for valve displacement \cite{fumagalli2021reduced, korakianitis2006numerical} and methods allowing the simulation of isovolumetric phases of the cardiac cycle \cite{zingaro2022modeling, this2020augmented}. Moreover, we plan on introducing more advanced methods for the domain displacement, including non-linear lifting operators \cite{bucelli2022mathematical, bakir2018multiphysics, alharbi2022fluid} and remeshing techniques \cite{lantz2021impact}. Finally, the inclusion of scalar transport models \cite{mittal2016computational, corti2022impact, seo2014effect} and non-Newtonian rheologies for blood \cite{de2016numerical, al2017investigating} will allow to better replicate several scenarios of clinical interest.

\FloatBarrier
\newpage

\section*{Author contributions}
\textbf{Pasquale Claudio Africa}: conceptualization, methodology, software (development and maintenance), formal analysis, project administration, writing (original draft).
\textbf{Ivan Fumagalli}: conceptualization, methodology, software (development and simulation), formal analysis, writing (original draft).
\textbf{Michele Bucelli}: conceptualization, methodology, software (development, simulation and maintenance), visualization, formal analysis, project administration, writing (original draft). 
\textbf{Alberto Zingaro}: conceptualization, methodology, software (development and simulation), visualization, formal analysis, writing (original draft).
\textbf{Marco Fedele}: methodology, software (development), writing (review and editing).
\textbf{Luca Dede'}: methodology, project administration, writing (review and editing), supervision.
\textbf{Alfio Quarteroni}: funding acquisition, project administration, writing (review and editing), supervision.

\section*{Acknowledgements}

This project has received funding from the European Research Council (ERC) under the European Union's Horizon 2020 research and innovation program (grant agreement No 740132, iHEART - An Integrated Heart Model for the simulation of the cardiac function, P.I. Prof. A. Quarteroni).
\begin{center}
	\raisebox{-.5\height}{\includegraphics[width=.15\textwidth]{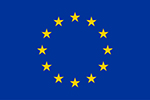}}
	\hspace{2cm}
	\raisebox{-.5\height}{\includegraphics[width=.15\textwidth]{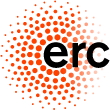}}
\end{center}

I.F. has also received support from ICSC--Centro Nazionale di Ricerca in High Performance Computing, Big Data, and Quantum Computing funded by the European Union's NextGenerationEU program.

The review process was conducted while P.C.A. was affiliated with SISSA, within the research activities of the consortium iNEST (Interconnected North-East Innovation Ecosystem), Piano Nazionale di Ripresa e Resilienza (PNRR) – Missione 4 Componente 2, Investimento 1.5 – D.D. 1058 23/06/2022, ECS00000043, supported by the European Union's NextGenerationEU program.

P.C.A. acknowledges the support from the Italian Ministry of University and Research (MUR), grant ``Dipartimento di Eccellenza 2023-2027'', Mathematics Area, SISSA.
I.F., M.B., L.D., and A.Q. acknowledge the support from the Italian Ministry of University and Research (MUR), grant ``Dipartimento di Eccellenza 2023-2027'', Department of Mathematics, Politecnico di Milano.

The authors acknowledge the CINECA award under the ISCRA B initiative for the availability of high performance computing resources and support (ISCRA grant IsB25\_MathBeat, P.I. Alfio Quarteroni, 2022--2023).

All the authors are members of the INdAM group GNCS ``Gruppo Nazionale per il Calcolo Scientifico'' (National Group for Scientific Computing).

\appendix

\section{Beltrami flow}
\label{app:beltrami}

For the sake of reproducibility, we report the details of the test case with analytical solution considered in \cref{sec:software-verification} and taken from \cite{ethier1994exact}. We consider Navier-Stokes equations with the \textit{gradient-gradient} formulation of the diffusion term (see \eqref{eq:diffusion-term}). The domain is the cube $\Omega = (0, 2\pi)^3$. The solution has the following analytical expression:
\begin{align*}
    u_x(\bm x, t) &= -\frac{c(t)}{4} \left[ \exp(\frac{\pi x}{4}) \sin\left(\frac{\pi(y + 2z)}{4}\right) + \exp(\frac{\pi z}{4})\cos(\frac{\pi(x + 2y)}{4}) \right]\;, \\[1em]
    u_y(\bm x, t) &= -\frac{c(t)}{4} \left[ \exp(\frac{\pi y}{4}) \sin\left(\frac{\pi(2x + z)}{4}\right) + \exp(\frac{\pi x}{4})\cos(\frac{\pi(y + 2z)}{4}) \right]\;, \\[1em]
    u_z(\bm x, t) &= -\frac{c(t)}{4} \left[ \exp(\frac{\pi z}{4}) \sin\left(\frac{\pi(x + 2y)}{4}\right) + \exp(\frac{\pi y}{4})\cos(\frac{\pi(2x + z)}{4}) \right]\;,
\end{align*}
\begin{align*}
    p(\bm x, t) = -\frac{\rho\pi^2}{32} \exp(-\frac{\pi^2 t\mu}{2\rho}) \bigg[ 
    & \exp(\frac{\pi x}{2}) + \exp(\frac{\pi y}{2}) + \exp(\frac{\pi z}{2}) \\
    & + 2\exp(\frac{\pi(x + y)}{4})\cos(\frac{\pi(y + 2z)}{4})\sin(\frac{\pi(2x + z)}{4}) \\
    & + 2\exp(\frac{\pi(x + z)}{4})\cos(\frac{\pi(x + 2y)}{4})\sin(\frac{\pi(y + 2z)}{4}) \\
    & + 2\exp(\frac{\pi(y + z)}{4})\cos(\frac{\pi(2x + z)}{4})\sin(\frac{\pi(x + 2y)}{4})
    \bigg]\;,
\end{align*}
wherein $\vel = [u_x, u_y, u_z]^T$, $\bm x = [x, y, z]^T$ and
\begin{equation*}
    c(t) = \pi\exp(-\frac{\pi^2\mu t}{4\rho})\;.
\end{equation*}

The above functions satisfy Navier-Stokes equations with null forcing term ($\bm f = \bm 0$) and no resistive surfaces.

We displace the domain by computing the harmonic extension of the following boundary displacement:
\begin{equation*}
    \displ_{\widehat{\Gamma}^\mathrm{D}}(\xhat, t) = \num{50}\,\sin(t)\,\sin(\xhat)\;,
\end{equation*}
with $\widehat{\Gamma}^\mathrm{D} = \partial\widehat{\Omega}$ and $\sin(\xhat) = (\sin(x),\;\sin(y),\;\sin(z))^T$. The boundary displacement is such that mesh nodes on boundary faces only displace tangentially to the boundary (i.e. the domain is the cube $(0, 2\pi)$ at all times). Therefore, the exact solution described above satisfies Navier-Stokes equations in the moving reference as well. Furthermore, the dependency of $\displ_{\widehat{\Gamma}^\mathrm{D}}$ on time is such that the maximum value of $\uale$ is comparable in magnitude to the maximum value of $\vel$.

We obtain the initial conditions by setting $t = 0$ in the above, and we set Dirichlet conditions on all sides of the cube except for the side $\{ x = 0, y \in (0, 2\pi),$ $z \in (0, 2\pi)\}$, where we impose Neumann conditions. All boundary conditions are derived from the exact solution. We set density $\rho = \SI{1}{\kilo\gram\per\cubic\metre}$ and viscosity $\mu = \SI{10}{\pascal\second}$.

\bibliographystyle{elsarticle-num}
\bibliography{main}{}

\begin{thebibliography}{100}
\expandafter\ifx\csname url\endcsname\relax
  \def\url#1{\texttt{#1}}\fi
\expandafter\ifx\csname urlprefix\endcsname\relax\def\urlprefix{URL }\fi
\expandafter\ifx\csname href\endcsname\relax
  \def\href#1#2{#2} \def\path#1{#1}\fi

\bibitem{formaggia2010cardiovascular}
L.~Formaggia, A.~Quarteroni, A.~Veneziani, Cardiovascular Mathematics: Modeling
  and simulation of the circulatory system, Vol.~1, Springer Science \&
  Business Media, 2010.

\bibitem{chnafa2016image}
C.~Chnafa, S.~Mendez, F.~Nicoud, Image-based simulations show important flow
  fluctuations in a normal left ventricle: what could be the implications?,
  Annals of Biomedical Engineering 44~(11) (2016) 3346--3358.

\bibitem{collia2019analysis}
D.~Collia, L.~Zovatto, G.~Pedrizzetti, Analysis of mitral valve regurgitation
  by computational fluid dynamics, APL Bioengineering 3~(3) (2019) 036105.

\bibitem{luraghi2018numerical}
G.~Luraghi, W.~Wu, H.~De~Castilla, J.~F. Rodriguez~Matas, G.~Dubini, P.~Dubuis,
  M.~Grimm{\'e}, F.~Migliavacca, Numerical approach to study the behavior of an
  artificial ventricle: fluid--structure interaction followed by fluid dynamics
  with moving boundaries, Artificial Organs 42~(10) (2018) E315--E324.

\bibitem{luraghi2020impact}
G.~Luraghi, J.~F.~R. Matas, M.~Beretta, N.~Chiozzi, L.~Iannetti,
  F.~Migliavacca, The impact of calcification patterns in transcatheter aortic
  valve performance: a fluid-structure interaction analysis, Computer Methods
  in Biomechanics and Biomedical Engineering 24~(4) (2020) 375--383.

\bibitem{goubergrits2022ct}
L.~Goubergrits, K.~Vellguth, L.~Obermeier, A.~Schlief, L.~Tautz, J.~Bruening,
  H.~Lamecker, A.~Szengel, O.~Nemchyna, C.~Knosalla, et~al., {CT}-based
  analysis of left ventricular hemodynamics using statistical shape modeling
  and computational fluid dynamics, Frontiers in Cardiovascular Medicine 9
  (2022).

\bibitem{karabelas2022global}
E.~Karabelas, S.~Longobardi, J.~Fuchsberger, O.~Razeghi, C.~Rodero,
  M.~Strocchi, R.~Rajani, G.~Haase, G.~Plank, S.~Niederer, Global sensitivity
  analysis of four chamber heart hemodynamics using surrogate models, IEEE
  Transactions on Biomedical Engineering (2022).

\bibitem{kronborg2022computational}
J.~Kronborg, F.~Svelander, S.~Eriksson-Lidbrink, L.~Lindstr{\"o}m,
  C.~Homs-Pons, D.~Lucor, J.~Hoffman, Computational analysis of flow structures
  in turbulent ventricular blood flow associated with mitral valve
  intervention, Frontiers in Physiology (2022) 752.

\bibitem{mittal2016computational}
R.~Mittal, J.~H. Seo, V.~Vedula, Y.~J. Choi, H.~Liu, H.~H. Huang, S.~Jain,
  L.~Younes, T.~Abraham, R.~T. George, Computational modeling of cardiac
  hemodynamics: current status and future outlook, Journal of Computational
  Physics 305 (2016) 1065--1082.

\bibitem{quarteroni2019mathematical}
A.~Quarteroni, L.~Dede', A.~Manzoni, C.~Vergara, Mathematical modelling of the
  human cardiovascular system: data, numerical approximation, clinical
  applications, Vol.~33, Cambridge University Press, 2019.

\bibitem{this2020pipeline}
A.~This, H.~G. Morales, O.~Bonnefous, M.~A. Fern{\'a}ndez, J.-F. Gerbeau, A
  pipeline for image based intracardiac {CFD} modeling and application to the
  evaluation of the {PISA} method, Computer Methods in Applied Mechanics and
  Engineering 358 (2020) 112627.

\bibitem{zingaro2021hemodynamics}
A.~Zingaro, L.~Dede', F.~Menghini, A.~Quarteroni, Hemodynamics of the heart’s
  left atrium based on a {Variational Multiscale-LES} numerical method,
  European Journal of Mechanics-B/Fluids 89 (2021) 380--400.

\bibitem{viola2020fluid}
F.~Viola, V.~Meschini, R.~Verzicco, {Fluid--Structure-Electrophysiology
  interaction (FSEI) in the left-heart: a multi-way coupled computational
  model}, European Journal of Mechanics-B/Fluids 79 (2020) 212--232.

\bibitem{viola2022fsei}
F.~Viola, V.~Spandan, V.~Meschini, J.~Romero, M.~Fatica, M.~D. de~Tullio,
  R.~Verzicco, {FSEI-GPU: GPU accelerated simulations of the
  fluid--structure--electrophysiology interaction in the left heart}, Computer
  physics communications 273 (2022) 108248.

\bibitem{santiago2018fully}
A.~Santiago, J.~Aguado-Sierra, M.~Zavala-Ak{\'e}, R.~Doste-Beltran,
  S.~G{\'o}mez, R.~Ar{\'\i}s, J.~C. Cajas, E.~Casoni, M.~V{\'a}zquez, Fully
  coupled fluid-electro-mechanical model of the human heart for supercomputers,
  International journal for numerical methods in biomedical engineering 34~(12)
  (2018) e3140.

\bibitem{viscardi2010comparative}
F.~Viscardi, C.~Vergara, L.~Antiga, S.~Merelli, A.~Veneziani, G.~Puppini,
  G.~Faggian, A.~Mazzucco, G.~B. Luciani, Comparative finite element model
  analysis of ascending aortic flow in bicuspid and tricuspid aortic valve,
  Artificial organs 34~(12) (2010) 1114--1120.

\bibitem{faggiano2013helical}
E.~Faggiano, L.~Antiga, G.~Puppini, A.~Quarteroni, G.~B. Luciani, C.~Vergara,
  Helical flows and asymmetry of blood jet in dilated ascending aorta with
  normally functioning bicuspid valve, Biomechanics and modeling in
  mechanobiology 12 (2013) 801--813.

\bibitem{tagliabue2017fluid}
A.~Tagliabue, L.~Dede, A.~Quarteroni, Fluid dynamics of an idealized left
  ventricle: the extended nitsche's method for the treatment of heart valves as
  mixed time varying boundary conditions, International Journal for Numerical
  Methods in Fluids 85~(3) (2017) 135--164.

\bibitem{tricerri2015fluid}
P.~Tricerri, L.~Ded{\`e}, S.~Deparis, A.~Quarteroni, A.~M. Robertson,
  A.~Sequeira, Fluid-structure interaction simulations of cerebral arteries
  modeled by isotropic and anisotropic constitutive laws, Computational
  Mechanics 55 (2015) 479--498.

\bibitem{fumagalli2023fluid}
I.~Fumagalli, R.~Polidori, F.~Renzi, L.~Fusini, A.~Quarteroni, G.~Pontone,
  C.~Vergara, Fluid-structure interaction analysis of transcatheter aortic
  valve implantation, International Journal for Numerical Methods in Biomedical
  Engineering (2023) e3704.

\bibitem{paliwal2021presence}
N.~Paliwal, R.~L. Ali, M.~Salvador, R.~O’Hara, R.~Yu, U.~A. Daimee,
  T.~Akhtar, P.~Pandey, D.~D. Spragg, H.~Calkins, et~al., Presence of left
  atrial fibrosis may contribute to aberrant hemodynamics and increased risk of
  stroke in atrial fibrillation patients, Frontiers in Physiology (2021) 684.

\bibitem{mill2021sensitivity}
J.~Mill, V.~Agudelo, A.~L. Olivares, M.~I. Pons, E.~Silva, M.~Nu{\~n}ez-Garcia,
  X.~Morales, D.~Arzamendi, X.~Freixa, J.~Noailly, et~al., Sensitivity analysis
  of in silico fluid simulations to predict thrombus formation after left
  atrial appendage occlusion, Mathematics 9~(18) (2021) 2304.

\bibitem{morales2021deep}
X.~Morales~Ferez, J.~Mill, K.~A. Juhl, C.~Acebes, X.~Iriart, B.~Legghe,
  H.~Cochet, O.~De~Backer, R.~R. Paulsen, O.~Camara, Deep learning framework
  for real-time estimation of in-silico thrombotic risk indices in the left
  atrial appendage, Frontiers in Physiology 12 (2021) 694945.

\bibitem{oks2022fluid}
D.~Oks, C.~Samaniego, G.~Houzeaux, C.~Butakoff, M.~V{\'a}zquez,
  Fluid--structure interaction analysis of eccentricity and leaflet rigidity on
  thrombosis biomarkers in bioprosthetic aortic valve replacements,
  International Journal for Numerical Methods in Biomedical Engineering 38~(12)
  (2022) e3649.

\bibitem{santiago2022design}
A.~Santiago, C.~Butakoff, B.~Eguzkitza, R.~A. Gray, K.~May-Newman,
  P.~Pathmanathan, V.~Vu, M.~V{\'a}zquez, Design and execution of a
  verification, validation, and uncertainty quantification plan for a numerical
  model of left ventricular flow after {LVAD} implantation, PLoS computational
  biology 18~(6) (2022) e1010141.

\bibitem{domenichini2005three}
F.~Domenichini, G.~Pedrizzetti, B.~Baccani, Three-dimensional filling flow into
  a model left ventricle, Journal of fluid mechanics 539 (2005) 179--198.

\bibitem{domenichini2007combined}
F.~Domenichini, G.~Querzoli, A.~Cenedese, G.~Pedrizzetti, Combined experimental
  and numerical analysis of the flow structure into the left ventricle, Journal
  of biomechanics 40~(9) (2007) 1988--1994.

\bibitem{seo2013effect}
J.~H. Seo, R.~Mittal, Effect of diastolic flow patterns on the function of the
  left ventricle, Physics of Fluids 25~(11) (2013) 110801.

\bibitem{seo2014effect}
J.~H. Seo, V.~Vedula, T.~Abraham, A.~C. Lardo, F.~Dawoud, H.~Luo, R.~Mittal,
  Effect of the mitral valve on diastolic flow patterns, Physics of fluids
  26~(12) (2014) 121901.

\bibitem{tagliabue2017complex}
A.~Tagliabue, L.~Dede', A.~Quarteroni, {Complex blood flow patterns in an
  idealized left ventricle: A numerical study}, Chaos 27~(9) (2017) 093939.

\bibitem{dede2021computational}
L.~Dede', F.~Menghini, A.~Quarteroni, Computational fluid dynamics of blood
  flow in an idealized left human heart, International Journal for Numerical
  Methods in Biomedical Engineering 37~(11) (2021) e3287.

\bibitem{masci2020proof}
A.~Masci, M.~Alessandrini, D.~Forti, F.~Menghini, L.~Dede', C.~Tomasi,
  A.~Quarteroni, C.~Corsi, A proof of concept for computational fluid dynamic
  analysis of the left atrium in atrial fibrillation on a patient-specific
  basis, Journal of Biomechanical Engineering 142~(1) (2020).

\bibitem{corti2022impact}
M.~Corti, A.~Zingaro, L.~Dede', A.~Quarteroni, Impact of atrial fibrillation on
  left atrium haemodynamics: A computational fluid dynamics study, Computers in
  Biology and Medicine (2022) 106143.

\bibitem{di2021computational}
S.~Di~Gregorio, M.~Fedele, G.~Pontone, A.~F. Corno, P.~Zunino, C.~Vergara,
  A.~Quarteroni, A computational model applied to myocardial perfusion in the
  human heart: from large coronaries to microvasculature, Journal of
  Computational Physics 424 (2021) 109836.

\bibitem{barnafi2022multiscale}
N.~A. Barnafi~Wittwer, S.~D. Gregorio, L.~Dede', P.~Zunino, C.~Vergara,
  A.~Quarteroni, A multiscale poromechanics model integrating myocardial
  perfusion and the epicardial coronary vessels, SIAM Journal on Applied
  Mathematics 82~(4) (2022) 1167--1193.

\bibitem{sacco2018left}
F.~Sacco, B.~Paun, O.~Lehmkuhl, T.~L. Iles, P.~A. Iaizzo, G.~Houzeaux,
  M.~V{\'a}zquez, C.~Butakoff, J.~Aguado-Sierra, Left ventricular
  trabeculations decrease the wall shear stress and increase the
  intra-ventricular pressure drop in {CFD} simulations, Frontiers in Physiology
  9 (2018) 458.

\bibitem{vedula2016effect}
V.~Vedula, J.-H. Seo, A.~C. Lardo, R.~Mittal, Effect of trabeculae and
  papillary muscles on the hemodynamics of the left ventricle, Theoretical and
  Computational Fluid Dynamics 30~(1) (2016) 3--21.

\bibitem{brown2023patient}
J.~A. Brown, J.~H. Lee, M.~A. Smith, D.~R. Wells, A.~Barrett, C.~Puelz, J.~P.
  Vavalle, B.~E. Griffith, Patient--specific immersed finite
  element--difference model of transcatheter aortic valve replacement, Annals
  of biomedical engineering 51~(1) (2023) 103--116.

\bibitem{kung2021hemodynamics}
E.~Kung, C.~Baker, C.~Corsini, A.~Baretta, G.~Biglino, G.~Arbia, S.~Pant,
  A.~Marsden, A.~Taylor, M.~Quail, I.~Vignon-Clementel, G.~Pennati,
  F.~Migliavacca, S.~Schievano, A.~Hlavacek, A.~Dorfman, T.-Y. Hsia,
  R.~Figliola, Hemodynamics after {Fontan} procedure are determined by patient
  characteristics and anastomosis placement not graft selection: a
  patient-specific multiscale computational study, medRxiv (2021).

\bibitem{lantz2021impact}
J.~Lantz, S.~B{\"a}ck, C.-J. Carlh{\"a}ll, A.~Bolger, A.~Persson, M.~Karlsson,
  T.~Ebbers, Impact of prosthetic mitral valve orientation on the ventricular
  flow field: Comparison using patient-specific computational fluid dynamics,
  Journal of Biomechanics 116 (2021) 110209.

\bibitem{rigatelli2021applications}
G.~Rigatelli, C.~Chiastra, G.~Pennati, G.~Dubini, F.~Migliavacca, M.~Zuin,
  Applications of computational fluid dynamics to congenital heart diseases: a
  practical review for cardiovascular professionals, Expert Review of
  Cardiovascular Therapy 19~(10) (2021) 907--916.

\bibitem{oks2023effect}
D.~Oks, G.~Houzeaux, M.~V{\'a}zquez, M.~Neidlin, C.~Samaniego, Effect of {TAVR}
  commissural alignment on coronary flow: a fluid-structure interaction
  analysis, Computer Methods and Programs in Biomedicine (2023) 107818.

\bibitem{timmis2020european}
A.~Timmis, N.~Townsend, C.~P. Gale, A.~Torbica, M.~Lettino, S.~E. Petersen,
  E.~A. Mossialos, A.~P. Maggioni, D.~Kazakiewicz, H.~T. May, et~al., {European
  Society of Cardiology: cardiovascular disease statistics 2019}, European
  Heart Journal 41~(1) (2020) 12--85.

\bibitem{virani2020heart}
S.~S. Virani, A.~Alonso, E.~J. Benjamin, M.~S. Bittencourt, C.~W. Callaway,
  A.~P. Carson, A.~M. Chamberlain, A.~R. Chang, S.~Cheng, F.~N. Delling,
  et~al., Heart disease and stroke statistics—2020 update: a report from the
  {A}merican {H}eart {A}ssociation, Circulation 141~(9) (2020) e139--e596.

\bibitem{updegrove2017simvascular}
A.~Updegrove, N.~M. Wilson, J.~Merkow, H.~Lan, A.~L. Marsden, S.~C. Shadden,
  {SimVascular}: an open source pipeline for cardiovascular simulation, Annals
  of Biomedical Engineering 45 (2017) 525--541.

\bibitem{de2016numerical}
F.~De~Vita, M.~De~Tullio, R.~Verzicco, {Numerical simulation of the
  non-Newtonian blood flow through a mechanical aortic valve}, Theoretical and
  Computational Fluid Dynamics 30~(1) (2016) 129--138.

\bibitem{marom2015numerical}
G.~Marom, Numerical methods for fluid--structure interaction models of aortic
  valves, Archives of Computational Methods in Engineering 22 (2015) 595--620.

\bibitem{spuhler2020high}
J.~H. Sp{\"u}hler, J.~Jansson, N.~Jansson, J.~Hoffman, A high performance
  computing framework for finite element simulation of blood flow in the left
  ventricle of the human heart, in: H.~Van~Brummelen, A.~Corsini, S.~Perotto,
  G.~Rozza (Eds.), Numerical Methods for Flows, Springer, 2020, pp. 155--164.

\bibitem{votta2013toward}
E.~Votta, T.~B. Le, M.~Stevanella, L.~Fusini, E.~G. Caiani, A.~Redaelli,
  F.~Sotiropoulos, Toward patient-specific simulations of cardiac valves:
  state-of-the-art and future directions, Journal of Biomechanics 46~(2) (2013)
  217--228.

\bibitem{chnafa2014image}
C.~Chnafa, S.~Mendez, F.~Nicoud, Image-based large-eddy simulation in a
  realistic left heart, Computers \& Fluids 94 (2014) 173--187.

\bibitem{nicoud2018large}
F.~Nicoud, C.~Chnafa, J.~Siguenza, V.~Zmijanovic, S.~Mendez, Large-eddy
  simulation of turbulence in cardiovascular flows, in: Biomedical Technology,
  Springer, 2018, pp. 147--167.

\bibitem{africa2022flexible}
P.~C. Africa, lifex: A flexible, high performance library for the numerical
  solution of complex finite element problems, SoftwareX 20 (2022) 101252.

\bibitem{dealII93}
D.~Arndt, W.~Bangerth, B.~Blais, M.~Fehling, R.~Gassm{\"o}ller, T.~Heister,
  L.~Heltai, U.~K{\"o}cher, M.~Kronbichler, M.~Maier, P.~Munch, J.-P. Pelteret,
  S.~Proell, K.~Simon, B.~Turcksin, D.~Wells, J.~Zhang, The \texttt{deal.II}
  library, version 9.3, Journal of Numerical Mathematics 29~(3) (2021)
  171--186.

\bibitem{beyond_cfd2023}
E.~L. Schwarz, L.~Pegolotti, M.~R. Pfaller, A.~L. Marsden, Beyond {CFD}:
  Emerging methodologies for predictive simulation in cardiovascular health and
  disease, Biophysics Reviews 4~(1) (2023) 011301.

\bibitem{maas2012febio}
S.~A. Maas, B.~J. Ellis, G.~A. Ateshian, J.~A. Weiss, {FEBio: finite elements
  for biomechanics}, Journal of biomechanical engineering 134~(1) (2012).

\bibitem{CHeart2016}
J.~Lee, A.~Cookson, I.~Roy, E.~Kerfoot, L.~Asner, G.~Vigueras, T.~Sochi,
  S.~Deparis, C.~Michler, N.~P. Smith, D.~A. Nordsletten, Multiphysics
  computational modeling in cheart, SIAM Journal on Scientific Computing 38~(3)
  (2016) C150--C178.

\bibitem{griffith2013ibamr}
B.~Griffith, A.~Bhalla, {IBAMR}: An adaptive and distributed-memory parallel
  implementation of the immersed boundary method,
  \url{https://ibamr.github.io/} (2013).

\bibitem{mortensen2015oasis}
M.~Mortensen, K.~Valen-Sendstad, {Oasis: a high-level/high-performance open
  source Navier--Stokes solver}, Computer physics communications 188 (2015)
  177--188.

\bibitem{ExaDG2020}
D.~Arndt, N.~Fehn, G.~Kanschat, K.~Kormann, M.~Kronbichler, P.~Munch, W.~A.
  Wall, J.~Witte, {ExaDG}: High-order discontinuous {G}alerkin for the
  {Exa-Scale}, in: H.-J. Bungartz, S.~Reiz, B.~Uekermann, P.~Neumann, W.~E.
  Nagel (Eds.), Software for Exascale Computing - SPPEXA 2016-2019, Springer
  International Publishing, Cham, 2020, pp. 189--224.

\bibitem{blais2020lethe}
B.~Blais, L.~Barbeau, V.~Bibeau, S.~Gauvin, T.~El~Geitani, S.~Golshan,
  R.~Kamble, G.~Mirakhori, J.~Chaouki, Lethe: An open-source parallel
  high-order adaptative {CFD} solver for incompressible flows, SoftwareX 12
  (2020) 100579.

\bibitem{chen2014openfoam}
G.~Chen, Q.~Xiong, P.~J. Morris, E.~G. Paterson, A.~Sergeev, Y.~Wang, Openfoam
  for computational fluid dynamics, Notices of the AMS 61~(4) (2014) 354--363.

\bibitem{brenneisen2021sequential}
J.~Brenneisen, A.~Daub, T.~Gerach, E.~Kovacheva, L.~Huetter, B.~Frohnapfel,
  O.~D{\"o}ssel, A.~Loewe, Sequential coupling shows minor effects of fluid
  dynamics on myocardial deformation in a realistic whole-heart model,
  Frontiers in Cardiovascular Medicine 8 (2021) 1967.

\bibitem{lyras2021comparison}
K.~G. Lyras, J.~Lee, Comparison of numerical implementations for modelling flow
  through arterial stenoses, International Journal of Mechanical Sciences 211
  (2021) 106780.

\bibitem{chen2023hemodynamic}
A.~Chen, A.~Azriff~Basri, N.~B. Ismail, K.~Arifin~Ahmad, Hemodynamic effects of
  subaortic stenosis on blood flow characteristics of a mechanical heart valve
  based on openfoam simulation, Bioengineering 10~(3) (2023) 312.

\bibitem{fenicsx2022}
M.~W. Scroggs, J.~S. Dokken, C.~N. Richardson, G.~N. Wells, Construction of
  arbitrary order finite element degree-of-freedom maps on polygonal and
  polyhedral cell meshes, ACM Trans. Math. Softw. 48~(2) (may 2022).

\bibitem{esmaily2013new}
M.~Esmaily-Moghadam, Y.~Bazilevs, A.~L. Marsden, A new preconditioning
  technique for implicitly coupled multidomain simulations with applications to
  hemodynamics, Computational Mechanics 52 (2013) 1141--1152.

\bibitem{esmaily2015bi}
M.~Esmaily-Moghadam, Y.~Bazilevs, A.~L. Marsden, A bi-partitioned iterative
  algorithm for solving linear systems arising from incompressible flow
  problems, Computer Methods in Applied Mechanics and Engineering 286 (2015)
  40--62.

\bibitem{trilinos-website}
{T}rilinos {P}roject {W}ebsite, \url{https://trilinos.github.io} (2023).

\bibitem{petsc-user-ref}
S.~Balay, S.~Abhyankar, M.~F. Adams, S.~Benson, J.~Brown, P.~Brune,
  K.~Buschelman, E.~Constantinescu, L.~Dalcin, A.~Dener, V.~Eijkhout,
  J.~Faibussowitsch, W.~D. Gropp, V.~Hapla, T.~Isaac, P.~Jolivet, D.~Karpeev,
  D.~Kaushik, M.~G. Knepley, F.~Kong, S.~Kruger, D.~A. May, L.~C. McInnes,
  R.~T. Mills, L.~Mitchell, T.~Munson, J.~E. Roman, K.~Rupp, P.~Sanan,
  J.~Sarich, B.~F. Smith, S.~Zampini, H.~Zhang, H.~Zhang, J.~Zhang, {PETSc/TAO}
  users manual, Tech. Rep. ANL-21/39 - Revision 3.18, Argonne National
  Laboratory (2022).

\bibitem{africa2023lifex_fiber}
P.~C. Africa, R.~Piersanti, M.~Fedele, L.~Dede', A.~Quarteroni, lifex-fiber: an
  open tool for myofibers generation in cardiac computational models, BMC
  Bioinformatics 24 (2023) 143.

\bibitem{africa2023matrix}
P.~C. Africa, M.~Salvador, P.~Gervasio, L.~Dede', A.~Quarteroni, A matrix--free
  high--order solver for the numerical solution of cardiac electrophysiology,
  Journal of Computational Physics 478 (2023) 111984.

\bibitem{cicci2022deep}
L.~Cicci, S.~Fresca, A.~Manzoni, Deep-hyromnet: A deep learning-based operator
  approximation for hyper-reduction of nonlinear parametrized pdes, Journal of
  Scientific Computing 93~(2) (2022) 57.

\bibitem{cicci2022efficient}
L.~Cicci, S.~Fresca, A.~Manzoni, A.~Quarteroni, Efficient approximation of
  cardiac mechanics through reduced order modeling with deep learning-based
  operator approximation, arXiv preprint arXiv:2202.03904 (2022).

\bibitem{cicci2022projection}
L.~Cicci, S.~Fresca, S.~Pagani, A.~Manzoni, A.~Quarteroni, et~al.,
  Projection-based reduced order models for parameterized nonlinear
  time-dependent problems arising in cardiac mechanics, Mathematics in
  Engineering 5~(2) (2022) 1--38.

\bibitem{fedele2023comprehensive}
M.~Fedele, R.~Piersanti, F.~Regazzoni, M.~Salvador, P.~C. Africa, M.~Bucelli,
  A.~Zingaro, A.~Quarteroni, et~al., A comprehensive and biophysically detailed
  computational model of the whole human heart electromechanics, Computer
  Methods in Applied Mechanics and Engineering 410 (2023) 115983.

\bibitem{piersanti20223d}
R.~Piersanti, F.~Regazzoni, M.~Salvador, A.~F. Corno, C.~Vergara,
  A.~Quarteroni, et~al., {3D--0D} closed-loop model for the simulation of
  cardiac biventricular electromechanics, Computer Methods in Applied Mechanics
  and Engineering 391 (2022) 114607.

\bibitem{regazzoni2022cardiac}
F.~Regazzoni, M.~Salvador, P.~Africa, M.~Fedele, L.~Dede', A.~Quarteroni, A
  cardiac electromechanical model coupled with a lumped-parameter model for
  closed-loop blood circulation, Journal of Computational Physics 457 (2022)
  111083.

\bibitem{salvador2021electromechanical}
M.~Salvador, M.~Fedele, P.~C. Africa, E.~Sung, A.~Prakosa, J.~Chrispin,
  N.~Trayanova, A.~Quarteroni, et~al., Electromechanical modeling of human
  ventricles with ischemic cardiomyopathy: numerical simulations in sinus
  rhythm and under arrhythmia, Computers in Biology and Medicine 136 (2021)
  104674.

\bibitem{zingaro2022geometric}
A.~Zingaro, I.~Fumagalli, L.~Dede', M.~Fedele, P.~C. Africa, A.~F. Corno, A.~M.
  Quarteroni, A geometric multiscale model for the numerical simulation of
  blood flow in the human left heart, Discrete and Continous Dynamical System -
  S 15~(8) (2022) 2391--2427.

\bibitem{zingaro2023electromechanics}
A.~Zingaro, M.~Bucelli, R.~Piersanti, F.~Regazzoni, L.~Dede', A.~Quarteroni, An
  electromechanics-driven fluid dynamics model for the simulation of the whole
  human heart, arXiv preprint arXiv:2301.02148 (2023).

\bibitem{fumagalli2022image}
I.~Fumagalli, P.~Vitullo, C.~Vergara, M.~Fedele, A.~F. Corno, S.~Ippolito,
  R.~Scrofani, A.~Quarteroni, Image-based computational hemodynamics analysis
  of systolic obstruction in hypertrophic cardiomyopathy, Frontiers in
  Physiology (2022) 2437.

\bibitem{marcinno2022computational}
F.~Marcinn\`{o}, A.~Zingaro, I.~Fumagalli, L.~Dede', C.~Vergara, A
  computational study of blood flow dynamics in the pulmonary arteries, Vietnam
  Journal of Mathematics (2022) 1--23.

\bibitem{bennati2023image}
L.~Bennati, C.~Vergara, V.~Giambruno, I.~Fumagalli, A.~F. Corno, A.~Quarteroni,
  G.~Puppini, G.~B. Luciani, An image-based computational fluid dynamics study
  of mitral regurgitation in presence of prolapse, Cardiovascular Engineering
  and Technology (2023) 1--19.

\bibitem{bennati2023turbulence}
L.~Bennati, V.~Giambruno, F.~Renzi, V.~Di~Nicola, C.~Maffeis, G.~Puppini, G.~B.
  Luciani, C.~Vergara, Turbulent blood dynamics in the left heart in the
  presence of mitral regurgitation: a computational study based on multi-series
  cine-{MRI}, Biomechanics and Modeling in Mechanobiology 22 (2023)
  1829–1846.

\bibitem{zingaro2022modeling}
A.~Zingaro, M.~Bucelli, I.~Fumagalli, L.~Dede', A.~Quarteroni, Modeling
  isovolumetric phases in cardiac flows by an augmented resistive immersed
  implicit surface method, International Journal for Numerical Methods in
  Biomedical Engineering (2023) e3767.

\bibitem{bucelli2022partitioned}
M.~Bucelli, L.~Dede', A.~Quarteroni, C.~Vergara, Partitioned and monolithic
  algorithms for the numerical solution of cardiac fluid-structure interaction,
  Communications in Computational Physics 32~(5) (2022) 1217--1256.

\bibitem{bucelli2022mathematical}
M.~Bucelli, A.~Zingaro, P.~C. Africa, I.~Fumagalli, L.~Dede', A.~M. Quarteroni,
  A mathematical model that integrates cardiac electrophysiology, mechanics and
  fluid dynamics: application to the human left heart, International Journal
  for Numerical Methods in Biomedical Engineering 39~(3) (2023) e3678.

\bibitem{bucelli2022stable}
M.~Bucelli, M.~G. Gabriel, A.~Quarteroni, G.~Gigante, C.~Vergara, A stable
  loosely-coupled scheme for cardiac electro-fluid-structure interaction,
  Journal of Computational Physics 490 (2023) 112326.

\bibitem{di2022prediction}
S.~Di~Gregorio, C.~Vergara, G.~M. Pelagi, A.~Baggiano, P.~Zunino, M.~Guglielmo,
  L.~Fusini, G.~Muscogiuri, A.~Rossi, M.~G. Rabbat, et~al., Prediction of
  myocardial blood flow under stress conditions by means of a computational
  model, European Journal of Nuclear Medicine and Molecular Imaging (2022)
  1--12.

\bibitem{zingaro2023comprehensive}
A.~Zingaro, C.~Vergara, L.~Dede, F.~Regazzoni, A.~Quarteroni, A comprehensive
  mathematical model for cardiac perfusion, Scientific Reports 13~(1) (2023)
  14220.

\bibitem{ethier1994exact}
C.~R. Ethier, D.~Steinman, Exact fully {3D Navier--Stokes} solutions for
  benchmarking, International Journal for Numerical Methods in Fluids 19~(5)
  (1994) 369--375.

\bibitem{perktold1994flow}
K.~Perktold, E.~Thurner, T.~Kenner, Flow and stress characteristics in rigid
  walled and compliant carotid artery bifurcation models, Medical \& biological
  engineering \& computing 32~(1) (1994) 19--26.

\bibitem{quarteroni2017cardiovascular}
A.~Quarteroni, A.~Manzoni, C.~Vergara, The cardiovascular system: Mathematical
  modelling, numerical algorithms and clinical applications, Vol.~26, Cambridge
  University Press, 2017.

\bibitem{taylor1996finite}
C.~A. Taylor, T.~J. Hughes, C.~K. Zarins, Finite element analysis of pulsatile
  flow in the abdominal aorta under resting and exercise conditions, in: ASME
  International Mechanical Engineering Congress and Exposition, Vol. 15403,
  American Society of Mechanical Engineers, 1996, pp. 81--82.

\bibitem{taylor1998finite}
C.~A. Taylor, T.~J. Hughes, C.~K. Zarins, Finite element modeling of blood flow
  in arteries, Computer methods in applied mechanics and engineering 158~(1-2)
  (1998) 155--196.

\bibitem{donea1982arbitrary}
J.~Donea, S.~Giuliani, J.-P. Halleux, An arbitrary {Lagrangian-Eulerian} finite
  element method for transient dynamic fluid-structure interactions, Computer
  methods in applied mechanics and engineering 33~(1-3) (1982) 689--723.

\bibitem{donea2004arbitrary}
J.~Donea, A.~Huerta, J.-P. Ponthot, A.~Rodriguez-Ferran, {A}rbitrary
  {L}agrangian–{E}ulerian {M}ethods, in: R.~d.~B. Erwin~Stein, T.~J. Hughes.
  (Eds.), Encyclopedia of Computational Mechanics, Vol. 1: Fundamentals, John
  Wiley \& Sons, 2004, Ch.~14, pp. 413--437.

\bibitem{hughes1981lagrangian}
T.~J.~R. Hughes, W.~K. Liu, T.~K. Zimmermann, {Lagrangian-Eulerian} finite
  element formulation for incompressible viscous flows, Computer Methods in
  Applied Mechanics and Engineering 29~(3) (1981) 329--349.

\bibitem{fedele2017patient}
M.~Fedele, E.~Faggiano, L.~Dede', A.~Quarteroni, A patient-specific aortic
  valve model based on moving resistive immersed implicit surfaces,
  Biomechanics and Modeling in Mechanobiology 16~(5) (2017) 1779--1803.

\bibitem{fumagalli2020image}
I.~Fumagalli, M.~Fedele, C.~Vergara, L.~Dede', S.~Ippolito, F.~Nicol{\`{o}},
  C.~Antona, R.~Scrofani, A.~Quarteroni, An image-based computational
  hemodynamics study of the systolic anterior motion of the mitral valve,
  Computers in Biology and Medicine 123 (2020) 103922.

\bibitem{stein2003mesh}
K.~Stein, T.~Tezduyar, R.~Benney, Mesh moving techniques for fluid-structure
  interactions with large displacements, J. Appl. Mech. 70~(1) (2003) 58--63.

\bibitem{jasak2006automatic}
H.~Jasak, Z.~Tukovic, Automatic mesh motion for the unstructured finite volume
  method, Transactions of FAMENA 30~(2) (2006) 1--20.

\bibitem{quarteroni2010numerical}
A.~Quarteroni, R.~Sacco, F.~Saleri, Numerical mathematics, Vol.~37, Springer
  Science \& Business Media, 2010.

\bibitem{kundu2015fluid}
P.~K. Kundu, I.~M. Cohen, D.~R. Dowling, Fluid mechanics, Academic press, 2015.

\bibitem{womersley1955method}
J.~R. Womersley, Method for the calculation of velocity, rate of flow and
  viscous drag in arteries when the pressure gradient is known, The Journal of
  physiology 127~(3) (1955) 553.

\bibitem{berselli2013exact}
L.~C. Berselli, P.~Miloro, A.~Menciassi, E.~Sinibaldi, Exact solution to the
  inverse {Womersley} problem for pulsatile flows in cylindrical vessels, with
  application to magnetic particle targeting, Applied Mathematics and
  Computation 219~(10) (2013) 5717--5729.

\bibitem{bertoglio2014tangential}
C.~Bertoglio, A.~Caiazzo, A tangential regularization method for backflow
  stabilization in hemodynamics, Journal of Computational Physics 261 (2014)
  162--171.

\bibitem{moghadam2011comparison}
M.~E. Moghadam, Y.~Bazilevs, T.-Y. Hsia, I.~E. Vignon-Clementel, A.~L. Marsden,
  A comparison of outlet boundary treatments for prevention of backflow
  divergence with relevance to blood flow simulations, Computational Mechanics
  48~(3) (2011) 277--291.

\bibitem{vignon2010outflow}
I.~E. Vignon-Clementel, C.~Figueroa, K.~Jansen, C.~Taylor, Outflow boundary
  conditions for {3D} simulations of non-periodic blood flow and pressure
  fields in deformable arteries, Computer Methods in Biomechanics and
  Biomedical Engineering 13~(5) (2010) 625--640.

\bibitem{bazilevs2009patient}
Y.~Bazilevs, J.~Gohean, T.~Hughes, R.~Moser, Y.~Zhang, Patient-specific
  isogeometric fluid--structure interaction analysis of thoracic aortic blood
  flow due to implantation of the {Jarvik 2000} left ventricular assist device,
  Computer Methods in Applied Mechanics and Engineering 198~(45-46) (2009)
  3534--3550.

\bibitem{bazilevs2008isogeometric}
Y.~Bazilevs, V.~M. Calo, T.~J. Hughes, Y.~Zhang, Isogeometric fluid-structure
  interaction: theory, algorithms, and computations, Computational mechanics 43
  (2008) 3--37.

\bibitem{quarteroni2016geometric}
A.~Quarteroni, A.~Veneziani, C.~Vergara, Geometric multiscale modeling of the
  cardiovascular system, between theory and practice, Computer Methods in
  Applied Mechanics and Engineering 302 (2016) 193--252.

\bibitem{curtiss1952integration}
C.~F. Curtiss, J.~O. Hirschfelder, Integration of stiff equations, Proceedings
  of the National Academy of Sciences of the United States of America 38~(3)
  (1952) 235.

\bibitem{forti2015semi}
D.~Forti, L.~Dede', Semi-implicit {BDF} time discretization of the
  {Navier--Stokes} equations with {VMS-LES} modeling in a high performance
  computing framework, Computers \& Fluids 117 (2015) 168--182.

\bibitem{tezduyar2003stabilization}
T.~Tezduyar, S.~Sathe, Stabilization parameters in {SUPG} and {PSPG}
  formulations, Journal of computational and applied mechanics 4~(1) (2003)
  71--88.

\bibitem{bazilevs2007variational}
Y.~Bazilevs, V.~Calo, J.~Cottrell, T.~Hughes, A.~Reali, G.~Scovazzi,
  Variational multiscale residual-based turbulence modeling for large eddy
  simulation of incompressible flows, Computer methods in applied mechanics and
  engineering 197~(1-4) (2007) 173--201.

\bibitem{takizawa2014st}
K.~Takizawa, Y.~Bazilevs, T.~E. Tezduyar, C.~C. Long, A.~L. Marsden,
  K.~Schjodt, {ST and ALE-VMS} methods for patient-specific cardiovascular
  fluid mechanics modeling, Mathematical Models and Methods in Applied Sciences
  24~(12) (2014) 2437--2486.

\bibitem{quarteroni2017numerical}
A.~Quarteroni, Numerical Models for Differential Problems, Vol.~16, Springer,
  2017.

\bibitem{saad2003iterative}
Y.~Saad, Iterative methods for sparse linear systems, SIAM, 2003.

\bibitem{xu2017algebraic}
J.~Xu, L.~Zikatanov, Algebraic multigrid methods, Acta Numerica 26 (2017)
  591--721.

\bibitem{quarteroni1999domain}
A.~Quarteroni, A.~Valli, Domain decomposition methods for partial differential
  equations, Oxford University Press, 1999.

\bibitem{deparis2014parallel}
S.~Deparis, G.~Grandperrin, A.~Quarteroni, Parallel preconditioners for the
  unsteady {Navier--Stokes} equations and applications to hemodynamics
  simulations, Computers \& Fluids 92 (2014) 253--273.

\bibitem{dealii2019design}
D.~Arndt, W.~Bangerth, D.~Davydov, T.~Heister, L.~Heltai, M.~Kronbichler,
  M.~Maier, J.-P. Pelteret, B.~Turcksin, D.~Wells, The {deal.II} finite element
  library: Design, features, and insights, Computers \& Mathematics with
  Applications 81 (2021) 407--422.

\bibitem{petsc-web-page}
S.~Balay, S.~Abhyankar, M.~F. Adams, S.~Benson, J.~Brown, P.~Brune,
  K.~Buschelman, E.~M. Constantinescu, L.~Dalcin, A.~Dener, V.~Eijkhout,
  J.~Faibussowitsch, W.~D. Gropp, V.~Hapla, T.~Isaac, P.~Jolivet, D.~Karpeev,
  D.~Kaushik, M.~G. Knepley, F.~Kong, S.~Kruger, D.~A. May, L.~C. McInnes,
  R.~T. Mills, L.~Mitchell, T.~Munson, J.~E. Roman, K.~Rupp, P.~Sanan,
  J.~Sarich, B.~F. Smith, S.~Zampini, H.~Zhang, H.~Zhang, J.~Zhang, {PETS}c
  {W}eb page, \url{https://petsc.org/} (2023).

\bibitem{schroeder2006visualization}
W.~Schroeder, K.~M. Martin, W.~E. Lorensen, The visualization toolkit, Kitware,
  2006.

\bibitem{schaling2011boost}
B.~Sch{\"a}ling, The boost C++ libraries, Boris Sch{\"a}ling, 2011.

\bibitem{sacado-website}
{SACADO} {P}roject {W}ebsite, \url{https://trilinos.github.io/sacado} (2023).

\bibitem{girault2012finite}
V.~Girault, P.-A. Raviart, Finite element methods for Navier-Stokes equations:
  theory and algorithms, Vol.~5, Springer Science \& Business Media, 2012.

\bibitem{ladisa2011computational}
J.~F.~J. LaDisa, R.~J. Dholakia, C.~A. Figueroa, I.~E. Vignon-Clementel, F.~P.
  Chan, M.~M. Samyn, J.~R. Cava, C.~A. Taylor, J.~A. Feinstein, Computational
  simulations demonstrate altered wall shear stress in aortic coarctation
  patients treated by resection with end-to-end anastomosis, Congenital heart
  disease 6~(5) (2011) 432--443.

\bibitem{wilson2013vascular}
N.~M. Wilson, A.~K. Ortiz, A.~B. Johnson, The vascular model repository: a
  public resource of medical imaging data and blood flow simulation results,
  Journal of medical devices 7~(4), \url{https://www.vascularmodel.com/}
  (2013).

\bibitem{vmtk}
L.~Antiga, M.~Piccinelli, L.~Botti, B.~Ene-Iordache, A.~Remuzzi, D.~A.
  Steinman, An image-based modeling framework for patient-specific
  computational hemodynamics, Medical {\&} Biological Engineering {\&}
  Computing 46~(11) (2008) 1097--1112.

\bibitem{niederer:geom}
C.~H. Roney, R.~Bendikas, F.~Pashakhanloo, C.~Corrado, E.~J. Vigmond, E.~R.
  McVeigh, N.~A. Trayanova, S.~A. Niederer, Constructing a human atrial fibre
  atlas, Annals of Biomedical Engineering 49 (2021) 233--250.

\bibitem{roney:geom}
C.~H. Roney, R.~Bendikas, F.~Pashakhanloo, C.~Corrado, E.~J. Vigmond, E.~R.
  McVeigh, N.~A. Trayanova, S.~A. Niederer, Constructing a human atrial fibre
  atlas, \url{https://doi.org/10.5281/zenodo.3764917} (2020).

\bibitem{fedele2021polygonal}
M.~Fedele, A.~M. Quarteroni, Polygonal surface processing and mesh generation
  tools for numerical simulations of the complete cardiac function,
  International Journal for Numerical Methods in Biomedical Engineering 37
  (2021) e3435.

\bibitem{MV:open}
A.~G. Tsakiris, D.~A. Gordon, R.~Padiyar, D.~Fréchette, Relation of mitral
  valve opening and closure to left atrial and ventricular pressure in the
  intact dog, Am. J. Physiol. 234 (1978) H146--H151.

\bibitem{MV:close}
A.~\v{S}malcelj, D.~G. Gibson, Relation between mitral valve closure and early
  systolic function of the left ventricle, Heart 53 (1985) 436--442.

\bibitem{CRAWFORD20011419}
M.~H. Crawford, C.~A. Roldan, Quantitative assessment of valve thickness in
  normal subjects by transesophageal echocardiography, The American Journal of
  Cardiology 87~(12) (2001) 1419--1423.

\bibitem{brachet1983small}
M.~E. Brachet, D.~I. Meiron, S.~A. Orszag, B.~Nickel, R.~H. Morf, U.~Frisch,
  Small-scale structure of the {T}aylor--{G}reen vortex, Journal of Fluid
  Mechanics 130 (1983) 411--452.

\bibitem{tgv-dns-data}
Reference data for {T}aylor--{G}reen vortex benchmark problem,
  \url{https://cfd.ku.edu/hiocfd/spectral_Re1600_512.gdiag} (2011, last
  visited: 16 November 2023).

\bibitem{fumagalli2021reduced}
I.~Fumagalli, A reduced {3D-0D FSI} model of the aortic valve including
  leaflets curvature, arXiv preprint arXiv:2106.00571 (2021).

\bibitem{renzi2023accurate}
F.~Renzi, C.~Vergara, M.~Fedele, V.~Giambruno, A.~Quarteroni, G.~Puppini, G.~B.
  Luciani, Accurate and efficient 3d reconstruction of right heart shape and
  motion from multi-series cine-{MRI}, bioRxiv (2023) 2023--06.

\bibitem{korakianitis2006numerical}
T.~Korakianitis, Y.~Shi, Numerical simulation of cardiovascular dynamics with
  healthy and diseased heart valves, Journal of Biomechanics 39~(11) (2006)
  1964--1982.

\bibitem{this2020augmented}
A.~This, L.~Boilevin-Kayl, M.~A. Fern{\'a}ndez, J.-F. Gerbeau, Augmented
  resistive immersed surfaces valve model for the simulation of cardiac
  hemodynamics with isovolumetric phases, International Journal for Numerical
  Methods in Biomedical Engineering 36~(3) (2020) e3223.

\bibitem{bakir2018multiphysics}
A.~A. Bakir, A.~Al~Abed, M.~C. Stevens, N.~H. Lovell, S.~Dokos, A multiphysics
  biventricular cardiac model: Simulations with a left-ventricular assist
  device, Frontiers in Physiology 9 (2018) 1259.

\bibitem{alharbi2022fluid}
Y.~Alharbi, A.~Al~Abed, A.~A. Bakir, N.~H. Lovell, D.~W. Muller, J.~Otton,
  S.~Dokos, Fluid structure computational model of simulating mitral valve
  motion in a contracting left ventricle, Computers in Biology and Medicine 148
  (2022) 105834.

\bibitem{al2017investigating}
M.~G. Al-Azawy, A.~Turan, A.~Revell, Investigating the impact of
  {non-Newtonian} blood models within a heart pump, International Journal for
  Numerical Methods in Biomedical Engineering 33~(1) (2017) e02780.

\end{thebibliography}

\end{document}